# Re-examination of the SiGe Raman spectra – Linear chain approximation and a*b initio* calculations

O. Pagès,[1,*] J. Souhabi,[1] V. J. B. Torres,[2] A.V. Postnikov,[1] and K.C. Rustagi[3]

[1] *LPMD – IJB, Université Paul Verlaine –Metz, 57078 Metz, France*

[2] *Departamento de Física and I3N, Universidade de Aveiro, Campus Santiago, 3810 – 193, Aveiro, Portugal*

[3] *Indian Institute of Technology, Powai, Mumbai, 400076, India*

## Abstract

We report a detailed assignment of various features observed in the Raman spectra of SiGe alloys along the linear chain approximation (LCA), based on remarkable intensity-interplays with composition known from the literature, but so far not fully exploited. Such assignment is independently supported by *ab initio* calculation of the *frequencie*s of bond-stretching modes taking place in different local environments defined at one-dimension (1D), for sake of consistency with the LCA. Fair contour modeling of the SiGe Raman spectra is eventually obtained via a so-called 1D-Cluster version of the phenomenological Percolation scheme, as originally developed for zincblende alloys, after *ab initio* calibration of the intrinsic Si-Si, Si-Ge and Ge-Ge Raman efficiencies. The 1D-Cluster scheme introduces a seven-oscillator [1×(Ge-Ge), 4×(Si-Ge), 2×(Si-Si)] Raman behavior for SiGe, which considerably deviates from the currently accepted six-oscillator [1×(Ge-Ge), 1×(Si-Ge), 4×(Si-Si)] one. Different numbers of Raman modes per bond are interpreted as different sensitivities to the local environment of the Ge-Ge (insensitive), Si-Si (sensitive to 1[st] neighbors) and Si-Ge (sensitive to 2[nd] neighbors) bond-stretchings. The as-obtained SiGe Percolation scheme is also compared with the current version for zincblende alloys, using GaAsP as a natural reference. A marked deviation is concerned with an inversion of the like phonon branches in each multiplet. This is attributed either to the considerable Si and Ge phonon dispersions (Si-Si doublet), or to a basic difference in the lattice relaxations of diamond and zincblende alloys (Si-Ge multiplet). The SiGe vs. GaAsP comparison is supported by *ab initio* calculation of the local lattice relaxation/dynamics related to prototype impurity motifs that are directly transposable to the two crystal structures.

---

[*] Author to whom correspondence should be addressed:
Electronic mail: pages@univ-metz.fr

## I. Introduction

Since the emergence of semiconductor alloys some six decades ago, $Si_{1-x}Ge_x$, the leading group-IV alloy, with diamond structure (cubic), is by far the one whose physical properties have attracted most attention, both experimentally and theoretically.[1] Probably the reason is that the current semiconductor technology being essentially based on Si, any possible way to play around the intrinsic physical properties of Si, i.e. by alloying in the present case, is potentially interesting and as such methodically scrutinized. Moreover, among all possible Si-related group IV (C, Si, Ge, Sn) binary alloys, $Si_{1-x}Ge_x$ seems to be the only exception of a random alloy that retains the diamond structure of the parents throughout the whole composition domain. Certainly this is because the contrast in bond lengths ($l$) and in bond-stretching ($\alpha$) / bond-bending ($\beta$) force constants of the constituent species remains rather moderate in SiGe ($\Delta l \sim 4\%$, $\Delta\alpha \sim 15\%$, $\Delta\beta \sim 12\%$), while it becomes prohibitive for the next SiC (~34%, ~58%, ~83%) and SiSn (~17%, ~41%, ~53%) candidates.[2,3] On the technical side, a current application of major importance is that relaxed epilayers of SiGe can serve as a substrate for strained epilayers of Si,[4,5] which are then used for high mobility devices (Refs. [6] and [7]).

In this work we are interested in the vibrational properties of $Si_{1-x}Ge_x$, as conveniently detected at the laboratory scale by Raman scattering. While this technique operates at the Brillouin zone centre (BZC), and as such restricts the analysis to long-wavelength optical phonons ($q\sim0$), these are detected with a very high resolution, sometimes better than ~1 cm$^{-1}$. This is less, by at least a factor two, than the typical width at half maximum of the Raman line of a pure Si crystal.[8] As such, Raman scattering is certainly the best technique when searching to access the detail of the phonon mode behavior of a complex system such as an alloy. For a direct insight into the $Si_{1-x}Ge_x$ Raman spectra, we reproduce in **Fig. 1** a representative set taken by Alonso and Winer with epitaxial layers covering well-spanned alloy compositions (x~0.25, 0.50 and 0.75, taken from Fig. 2 of Ref. [9]).

Three main Ge-Ge, Si-Ge and Si-Si Raman features, underscored in **Fig. 1**, emerge at intermediate composition (x~0.55). The low (~300 cm$^{-1}$) and high (~500 cm$^{-1}$) frequency features connect to the modes of the pure Si and Ge crystals, respectively. Their assignment in terms of the main Ge-Ge and Si-Si modes is thus straightforward. The intermediate mode is intimately related to a mode observed in Si/Ge superlattices and in multiple quantum well structures when the Si/Ge interface is rough.[10,11] This naturally leads to an assignment in terms of the main Si-Ge mode.

The main Ge-Ge and Si-Si modes do exhibit opposite shifts when the Ge content increases, i.e. upward and downward, respectively, while the intermediate Si-Ge mode remains more or less stationary. A detailed modeling of the composition dependence of the frequencies of the main Ge-Ge, Si-

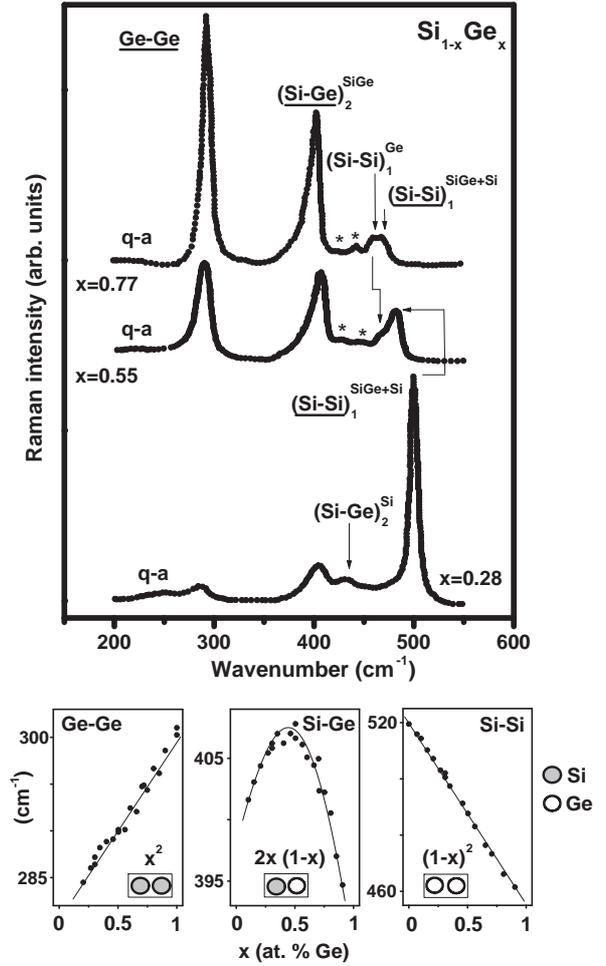

**Fig. 1**: **Main SiGe Raman features**. Representative $Si_{1-x}Ge_x$ Raman spectra taken by Alonso and Winer (plain symbols, digitalized from Fig. 2 of Ref. [9]). The x-dependencies proposed by Pezzoli et al. for the frequency (plain symbols, digitalized from Fig. 2 of ref. [13], guidelines are added for the eye) and intensity (in reference to the fractions of related oscillators) of the main Ge-Ge, Si-Ge and Si-Si Raman features (underscored) are summarized into specific panels. Each of these is assigned a specific bond-stretching oscillator (refer to the symbolic notation in each panel). We introduce a more general labeling of the Raman features, covering both the main (underscored) and the minor ones, corresponding to a given bond-stretching (main label) in a given 1D-environment. The latter is characterized by its microstructure (subscript: Ge-, SiGe- or Si-like) and length scale (superscript: 1$^{st}$- or 2$^{nd}$-neighbors). The corresponding oscillators are represented in the body of **Fig. 4**, using the same symbolic notation as in the present figure, for a direct comparison. Note that the minor $(Si-Ge)_2^{Si}$ feature decomposes into a pseudo-doublet away from the Ge-dilute/moderate limit (refer to the stars). The 'q-a' notation refers to a parasitical quasi-amorphous feature.

Ge and Si-Si Raman features was performed by Sui et al. using a modified cellular isodisplacement model.[12] The opposite Ge-Ge and Si-Si shifts were investigated by Rücker and Methfessel (Ref. [3]) by using a supercell approach and an anharmonic version of the Keating model, in which changes in the bond-stretching and bond-bending force constants depending on the local bond distortion are taken into account. They were attributed to a systematic screening of the microscopic strain effect by the confinement effect. Such competition is further discussed in this work at a later stage. As for the intensities of the main Ge-Ge, Si-Ge and Si-Si Raman lines, Pezzoli et al.[13] proposed empirical laws indicating a basic scaling with the bond fractions when assuming a random Si↔Ge substitution, i.e. according to $(1-x)^2$, $2x \cdot (1-x)$ and $x^2$, respectively. Incidentally, such scaling implies that the bonds of a given species can be considered as equivalent at a given alloy composition, i.e. vibrate at the same frequency whatever their local environment. This comes to assign the stretching of a chemical bond as the elementary oscillator behind each of the main Raman line. Such frequency (plain symbols, taken from Fig. 2 of Ref. 13) and intensity (see text and also Fig. 3 of Ref. 13) aspects are summarized per mode in specific panels at the bottom of Fig. 1.

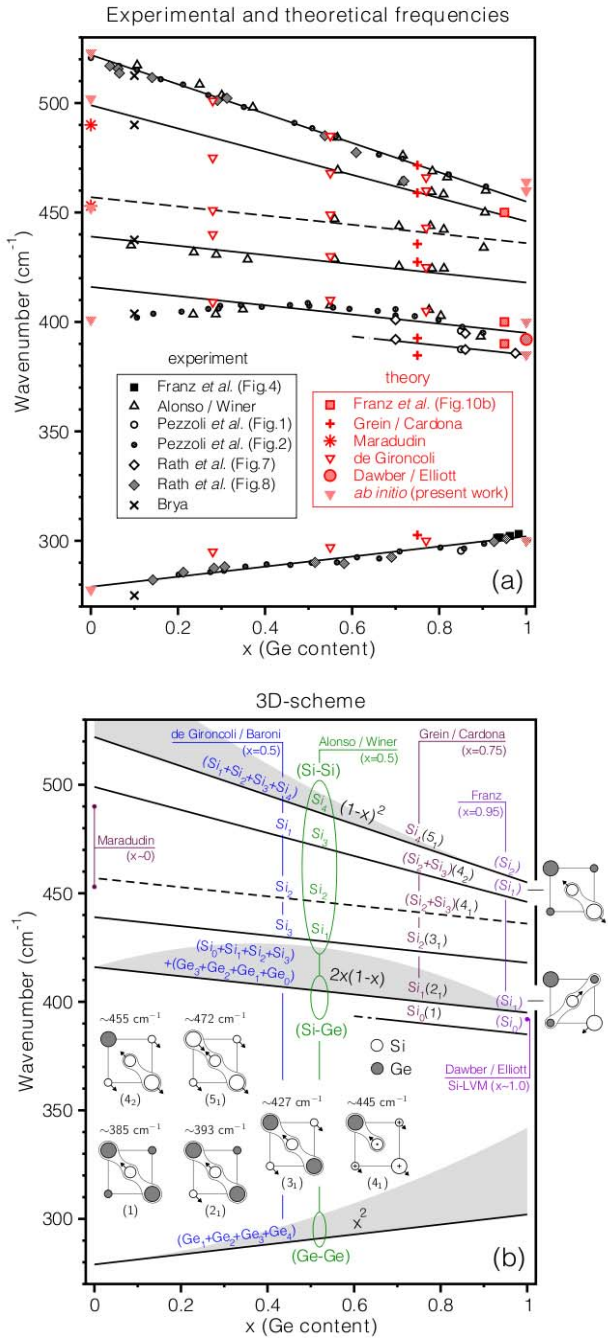

Fig. 2: **SiGe Raman spectra – State of the art**. $Si_{1-x}Ge_x$ Raman-frequency map including raw experimental (dark symbols, the cited data are from Refs. 9, 13, 15–17) and theoretical (clear symbols, the cited data are from Refs. 17–21) data from the literature [panel (a), straight guidelines are added for the eye], and corresponding assignment of the Raman modes within the current six-oscillator [1×(Ge-Ge), 1×(Si-Ge), 4×(Si-Si)] 3D-scheme [panel (b)]. The like phonon branches, referring to the same bond-stretching, are regrouped via ovals. The individual assignments are recapitulated by adopting a uniform $A_n$ notation. This refers to an atom A (Si or Ge) vibrating in presence of n A-like atoms (n=0-4) in its nearest-neighbor shell. The corresponding vibration patterns, as identified by Grein and Cardona within their mass-defect Green's function approach,[20] are schematically reproduced, in the body of panel (b), emphasizing bond-stretching. A standard notation is used for backward (⊕) and upward (⊙) atomic motions. The $N_M$ labeling of these authors, referring to the $M^{th}$ vibration mode in the sense of increasing frequency of a cluster of N Si atoms, is preserved, for a direct correspondence with the original patterns (given in Fig. 4 of Ref. 20). The individual bond fractions, which seem to monitor the intensities of the main Raman lines, following Pezzoli et al.,[13] are added for sake of completeness. They are specified on top of each relevant branch, and visualized via an appropriate thickening of such branches (grey area).

Now, careful examination of the SiGe Raman spectra shown in Fig. 1 reveals two additional minor features (not underscored) besides the main Ge-Ge, Si-Ge and Si-Si ones (underscored). An additional minor feature is discussed at a later stage. The assignment of the minor features has attracted considerable attention over the last four decades, both experimentally[9,14-17] and theoretically,[9,17-23] continuing so far. Some representative data are regrouped into a composite frequency map in Fig. 2a, their assignments being recapitulated in Fig. 2b. We emphasize that such assignments were systematically achieved based on sophisticated phonon calculations as performed in the real three-dimensional (3D) crystal, searching for *frequency*-matching between a given Raman line and the vibration of a given atom (Si or Ge) taken in some likely 3D environment at the considered alloy composition. In retrospect it seems that is was tacitly admitted that any 3D-environment should be limited to $1^{st}$-neighbors only. Accordingly we adopt a uniform $A_n$ notation for all 3D-assignments in Fig. 2b, corresponding to an atom A (Si or Ge) vibrating in presence of n (n=0–4) $1^{st}$-neighbors of the like species (Si or Ge, respectively). The additional information about the intensities of the main Ge-Ge, Si-Ge and Si-Si Raman lines, as empirically derived by Pezzoli et al. (see above),[13] is also included in Fig. 2b, for sake of completeness, via an appropriate thickening of the relevant phonon branches. As such, one may say that Fig. 2b summarizes the state of the art regarding the understanding of the SiGe Raman spectra.

The pioneering $Si_{1-x}Ge_x$ Raman spectra were obtained with polycrystals by Feldman et al.[14] at moderate Si content (0.75≤x≤1), and by Renucci et al.[24] and Brya[15] throughout the whole composition domain. Several interesting Raman features were identified. Preliminary indication regarding the origin of the emerging features at the Si- and Ge-dilute limits was given by Dawber and Elliott (Ref. 18), by Montroll and Potts (Ref. 23) and by Maradudin[19] using theories applicable to isolated impurities only. In particular, a long-standing enigma, as originally raised up by Brya[15] and taken up again by Taylor[25] in his review on the infrared and Raman properties of alloys, is that, quoting Brya,[15] '*the $402 cm^{-1}$ (main Si-Ge) line shows no obvious correspondence with any of the predicted modes … for an isolated Ge atom*', in reference to the two potential candidates predicted by Maradudin[19] (see Fig. 2) for the local vibration modes (LVM) of Ge in Si (Ge-LVM, $Ge_0$). At intermediate composition, where the existing theories could not apply, three distinct minor features were identified in between the main Si-Ge and Si-Si Raman features. They were tentatively assigned by Brya[15] as reminiscences of modes already detected in the Si- (~448 cm$^{-1}$) and Ge-dilute (~437 and ~487 cm$^{-1}$) limits, which had shown little change in frequency or intensity when the composition changes.

The next generation of data/assignments brings together the considerable experimental[9] and theoretical[9,20-22] efforts conducted before the mid-nineties in order to understand the multi-mode Raman pattern at intermediate composition, '*where little* (was) *known about the nature of the vibrational excitations*' quoting Brya (Ref. [15]). Such studies did benefit both from the arrival of a novel generation of samples, grown as monocrystals, and from the emergence of novel theoretical methods, that could treat the whole statistics of the alloy disorder on a realistic basis. Detail is given below.

Alonso and Winer (Ref. [9]) reported the first detailed Raman study covering a wide range of intermediate alloy compositions, the interpretation of their data being supported by a Keating model supercell (216-atom) calculation. All force constants were taken as constants in a first approximation. They were estimated from the elastic constants of the pure Si and Ge crystals,[2] taking geometrical means for the Si-Ge mixed bond. Reference calculations were performed with an ordered $Si_{0.5}Ge_{0.5}$ crystal corresponding to alternate pairs of Si and Ge (111) planes. The crucial result is that only the disordered supercells generated the experimentally observed minor features in between the main Si-Ge and Si-Si features. The trend is confirmed by an experimental observation that the minor features show up with bulk polycrystals,[9,14,15] as well as with relaxed epitaxial layers,[9,13,16,17] i.e. independently of the growth process. The origin of the minor features was further discussed at the microscopic scale by focusing onto the central minor peak at intermediate composition (x=0.55). This was attributed to Si-Si bond-stretching in a Si-centered tetrahedron unit with equal numbers (2) of Si and Ge atoms at the vertices ($Si_2$, ~450 cm$^{-1}$). The softening with respect to Si-Si stretching in the pure Si crystal ($Si_4$, ~520 cm$^{-1}$), corresponding to the main Si-Si Raman feature in the alloy ($Si_4$, ~480 cm$^{-1}$), was attributed to a confinement effect due to the quasi inert Ge atoms with a large mass. By extension, the three remaining minor peaks were assigned, in the sense of decreasing frequency, to more and a more confined Si-Si bond-stretching in Si-centered tetrahedron units with increasing number of Ge atoms at the vertices, i.e. from one ($Si_3$, ~470 cm$^{-1}$) to three ($Si_1$, ~430 cm$^{-1}$). *Altogether this yields the standard description of the SiGe Raman pattern with six bond-stretching oscillators [1×(Ge-Ge), 1×(Si-Ge), 4×(Si-Si)]*, except that the order of the (Si-Si) modes may vary from one author to another, as described below. The oscillators of like nature, i.e. referring to the same ultimate bond-stretching, are regrouped by ovals in Fig. 2b.

Grein and Cardona (Ref. [20]) used a mass-defect Green's-function approach to calculate the SiGe Raman spectra at moderate Si-content (25 at.%), at which limit the minor Raman features show up clearly. The alloy was ideally described in terms of a collection of Si- and Ge-centered tetrahedron units, each of these being treated as embedded in a uniform continuum defined along the virtual crystal approximation (VCA). Again, changes in the bond force constants due to local bond distortions, referred to as anharmonic effects, were neglected. The final Raman pattern at a given alloy composition was eventually reconstructed by weighting the Raman signals from the different tetrahedra by their occurrence in the crystal, assuming a random Si↔Ge distribution. A rather dense fine structure showed up above the main Si-Ge band, as expected. A common feature with previous calculations of Alonso and Winer (Ref. [9]) is that, globally, the Si-centered units with increasing number of Ge atoms at the vertices tend to generate Raman signals at lower frequencies. The vibration patterns of the modes that dominantly contribute to the minor Raman features were successively identified as 1 ($Si_0$, ~385 cm$^{-1}$), $2_1$ ($Si_1$, ~393 cm$^{-1}$), $3_1$ ($Si_2$, ~427 cm$^{-1}$), $4_1$ ($Si_3$, ~445 cm$^{-1}$), $4_2$ ($Si_3$, ~459 cm$^{-1}$), and $5_1$ ($Si_4$, ~471cm$^{-1}$). The corresponding vibration patterns are schematically reproduced in Fig. 2b. In doing so, we preserve the original notation $N_M$ of Grein and Cardona (within brackets),[20] for a direct reference to their work. The main label indicates the number of Si atoms per Si-centered unit, the subscript referring to increasing frequencies of the distinct vibration modes originating from that cluster. An important result is that a given peak among the minor features is usually due to different units, which contradicts an implicit assumption of Alonso and Winer (Ref. [9]) that there exists a univocal peak↔unit correspondence.

De Gironcoli (Ref. [21]) reported a full *ab initio* calculation of the Raman spectra of the disordered SiGe alloy using large supercells (512-atoms). Quoting the author, an '*almost perfect agreement with experiment* (was obtained) *for the weak structures*'.[21] Generally, this is consistent with an earlier conclusion of Alonso and Winer (Ref. [9]) that the weak structure in question are due to local fluctuations of the atomic structure in a disordered alloy.

De Gironcoli and Baroni (Ref. [22]) did provide further insight into the origin of the minor features by focusing their attention on the disordered $Si_{0.5}Ge_{0.5}$ alloy, and by tracing the overall Raman signal into a sequence of Ge- and Si-related partial phonon density of states being due to all possible Ge- and Si-centered tetrahedron units. In particular, the main Ge-Ge (at~300 cm$^{-1}$), Si-Ge (at ~400 cm$^{-1}$) and Si-Si (at~480 cm$^{-1}$) Raman features appear to be due to those tetrahedral units containing at least two Ge atoms ($Ge_1$-to-$Ge_4$), a mix of Si and Ge atoms ($Ge_0$-to-$Ge_3$ and $Si_0$-to-$Si_3$), and at least two Si atoms ($Si_1$-to-$Si_4$), respectively. Interestingly, the *ab initio* calculations reveal that the Ge atoms do not vibrate in the spectral range covered by the minor features. Therefore, these are all due to Si vibrations in different local (Si,Ge)- mixed environments, as anticipated by Alonso and Winer.[9] Now, de Gironcoli and Baroni (Ref. 22) found a basic correspondence between the minor features and the $Si_n$ units that is just opposite to that proposed by Alonso and Winer,[9] the large n values referring mainly to low frequency modes, and vice versa.

Summarizing, what emerges around the mid-nineties is that six oscillators [1×(Ge-Ge), 1×(Si-Ge), 4×(Si-Si)] would be sufficient to catch the SiGe Raman pattern in a nutshell (refer to the ovals in **Fig. 2b**), as originally proposed by Alonso and Winer (Ref. **9**). The high-frequency Si-Si oscillator ($Si_4$) evolves into the main Si-Si feature when the Si content increases, while the three remaining Si-Si oscillators ($Si_{1-2-3}$) remain minor and exhibit little change in frequency or intensity when the alloy composition changes, consistently with an original observation by Brya (see above, Ref. **15**). Grein and Cardona on the one hand,[20] and de Gironcoli and Baroni on the other hand,[21,22] did develop complementary phonon calculations at intermediate composition which have lead to refine the original assignment of Alonso and Winer,[9] without challenging the basic picture though.

Complications arise when entering the Si- and Ge-moderate/dilute limits. Recent experimental/theoretical insights, gained by Rath et al.[16] and by Franz et al.[17], reveal unsuspected trends therein. In particular a series of remarkable intensity-interplays, referred to as $RI_{1-2-3}$ hereafter, was detected in between neighboring Raman features. These play an important role in our work.

In the Si-Si spectral range, one such RI occurs at moderate Si content in between the main Si-Si mode (underscored) and the minor mode that grows on its low frequency side (not underscored), as apparent in the explicit data of Franz et al. (digitalized from Fig. 6b of Ref. **17**), reproduced in **Fig. 3c**. Such interplay was briefly mentioned by Alonso and Winer,[9] and is also observable in the Raman spectra reported by Rath et al. (refer to the higher frequency doublet at x=0.536, 0.70 and 0.86 in Fig. 11 of Ref. **16**). In fact, quasi-perfect intensity matching is detected around the critical Ge content $x_{c1}$~0.70-0.80 ($RI_1$). This is visible in **Fig. 1**, as well as in the corresponding theoretical spectra reported by de Gironcoli (refer to Fig. 5 of Ref. **21**). At lower Si content, the low-frequency peak becomes dominant. Eventually, in the Si-dilute limit, only that mode survives (see **Fig. 3c**). Based on their calculations using an anharmonic version of the Keating model and disordered large-size supercells (512-atom), Franz et al.[17] did assign the low- and high-frequency components of the doublet as being due to Si-Si bond-stretching, the Si-Si bond being either isolated in Ge ($Si_1$) or vibrating in presence of a third Si atom ($Si_2$), respectively (see **Fig. 2b**).

Somewhat surprisingly, similar scenarios develop for the main Si-Ge mode when entering the Si- and Ge-dilute limits, though on more restricted composition domains. This is apparent in the Raman data of both Rath et al. (taken from Fig. 1 of Ref. **16**) and Franz et al. (taken from Fig. 6b of Ref. **17**), reproduced in **Figs. 3b** and **3a**, respectively. In each case the main Si-Ge mode (underscored) is progressively relayed by a minor mode that grows on its high (x~0, see **Fig. 3b**) or low (x~1, see **Fig. 3a**) frequency side (not underscored). At a certain stage the side mode turns dominant. Remarkably, quasi-perfect intensity matching between the main and side features is achieved at nearly symmetrical contents of the minor species at the Si- ($x_{c2}$~0.9, $RI_2$ – see **Fig. 3a**) and Ge-dilute ($x_{c3}$~0.1, $RI_3$ – see **Fig. 3b**) limits. Explicit modelings with this respect are reported by Rath et al. (refer to Fig. 9/left of Ref. **16**, top and bottom panels).

At small Si content (x~1) the abundant data indicate that, eventually, the side mode at low frequency remains alone at high Si dilution, before total disappearance (see **Fig. 3a**). The side mode is thus naturally identified as the Si-LVM ($Si_0$). A reminiscence of the so-called main Si-Ge feature at high frequency just before total disappearance was attributed by Franz et al.[17] to the bending mode of an isolated Si-Si pair ($Si_1$). Note that such assignment of the nearby $Si_0$- and $Si_1$-like features around 400 cm$^{-1}$ was already achieved by Grein and Cardona (compare Fig. 10b of Ref. **17** with Fig. 3 of Ref. **20**). The intensity-interplay between the two features was not commented though, not even mentioned. A basic problem is that $RI_2$ brings in an additional oscillator (outside the ovals in **Fig. 2b**) on top of the six ones already identified by Alonso and Winer,[9] coming to a total of seven oscillators. Another problem is that the so-called main Si-Ge feature does not connect with the Si-LVM. This strongly reminds of the problem faced by Brya in the Ge-dilute limit, referring to the long-standing enigma.[15] For sake of completeness we mention that weak features were observed by Franz et al. on the low frequency side of the Si-LVM when reaching the highly-dilute limits, i.e. at x~0.99 (see Fig. 5 of Ref. **17**). However, these do not seem to be alloy-related, and were attributed by Franz et al.[17] to the natural Si isotopes in the crystal.

At small Ge content (x~0) the data are rather scarce. We note that the side mode at high frequency already dominates the so-called main Si-Ge mode at x=0.11, if we refer to the Raman spectrum taken by Rücker and Methfessel using the $Si_{0.878}Ge_{0.11}C_{0.012}$ alloy with quasi-negligible C content (refer to Fig. 6 of Ref. **3**). The trend is reinforced at x=0.07 (refer to **Fig. 3b**). It would have been interesting to see what happens to the doublet when approaching the highly-dilute Ge-limit, but we are not aware of any Raman data ever collected at a lower Ge content than 7 at.%. By analogy with $RI_2$ we anticipate that only the side mode at high-frequency will survive, eventually connecting with the Ge-LVM ($Ge_0$). This would solve the above-mentioned long-standing enigma raised up by Brya.[15]

In summary, at the term of this brief overview of the literature over the last four decades, certainly not an exhaustive one though, we are left with seven oscillators, and not only six as is usually admitted (refer to ovals in **Fig. 2b**), plus one basic problem regarding the assignment of the main Si-Ge feature, in reference to its non-convergence onto the Ge and Si LVM's, plus three interesting intensity-interplays as involved with the main Si-Si and Si-Ge features, in reference to **RI₁** and **RI₂₋₃**, respectively, plus one remarkable feature that the critical x values for RI₂ and RI₃ are nearly symmetrical (corresponding to ~10 impurity-at.%).

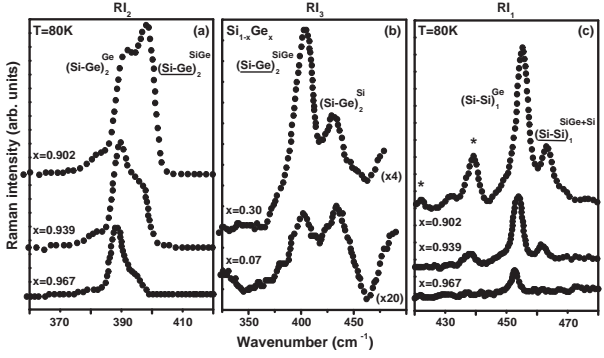

**Fig. 3**: **Remarkable intensity-interplays (RI's) between the main and minor SiGe Raman features**. $Si_{1-x}Ge_x$ Raman spectra representative of the **RI**'s taking place in between the main (underscored) and minor (not underscored) Raman features in the Si-Ge [panel (a) – **RI₂**, panel (b) – **RI₃**] and Si-Si [panel (c) – **RI₁**] spectral ranges when entering the Ge- [panel (b)] and Si-moderate/dilute [panels (a) and (c)] limits. The presented data in panels (a), (b) and (c) were digitalized from raw experimental spectra taken by Franz et al. at low temperature [panels (a), (c)], for a better resolution of neighboring Raman features, and by Rath et al., i.e. from Fig. 7-bottom of Ref. **17**, from Fig. 1 of Ref. **16** and from Fig. 6b of Ref. **17**, respectively. In panel (b) the data were reproduced after proper renormalization using the original factors, as specified within brackets. In panel (c) the stars refer to a decomposition of the original $(Si-Ge)_2^{Si}$ band into a pseudo-doublet (an overview is given in **Fig. 1**).

To our view, the latter point ($x_{c2} \sim 1-x_{c3}$), which has attracted no attention so far, cannot be merely fortuitous, and appeals a strictly similar assignment of the two side features within the same global scheme. This cannot be achieved within the current six-oscillator scheme reported in **Fig. 2b**, already because RI₂ involves a seventh oscillator. More generally, even if we forget about the problem of the seventh oscillator, any formalization of the observed **RI's** seems forbidden if one starts from the current 3D-assignment of the individual Raman lines in terms of the $Si_n$ and $Ge_n$ units (see **Fig. 2b**). This is because each unit may contribute to several Raman lines, and not to a single one.[20-22] This makes it extremely difficult, in principle, to derive explicit laws for the composition-dependence of the intensities of the individual Raman lines. In fact such explicit laws are still missing.

The main objective of the present work is to design a simple scheme that provides a consistent assignment of the whole set of Raman features over the whole composition domain, and, which, at the same time, naturally accounts for the observed composition dependence of the intensities of various modes, in reference to **RI₁₋₂₋₃**.

For doing so, we take advantage of a specificity of Raman scattering that is to operate at the BZC (q~0). Such restriction is interesting on the theoretical side, since at this limit the space phase term ($\vec{q} \cdot \vec{r}$) of the plane wave that ideally describes a phonon disappears, and with it all the information on the actual position ($\vec{r}$) of an atom in the real (3D) crystal. Therefore a phenomenological description of the lattice dynamics at one-dimension (1D), along the so-called linear chain approximation (LCA), should be sufficient in principle. In this case, the Raman active mode of a pure diamond crystal, say Si, which corresponds to out-of-phase displacements of the two intercalated *fcc* sub-lattices in the 3D diamond lattice, each of these being considered as quasi rigid (q~0), simply transposes at 1D into Si-Si bond-stretching. This generalizes to the SiGe alloy, meaning that all related Raman lines should be ultimately discussed in terms of Ge-Ge, Si-Ge and Si-Si bond-stretching at 1D. In principle, distinct Raman features may proceed from the same bond, which comes to distinguish between different local LCA-type (1D) environments of that bond then. Such 3D→1D change of scope for the very basic understanding of the 1-bond→multi-mode SiGe Raman spectra remains unexplored so far.

Three phenomenological, i.e. LCA-based (1D), schemes are available in the literature for the description of the Raman spectra of alloys, as originally worked out for $A_{1-x}B_xC$ zincblende crystals (C denoting indifferently a cation or an anion).

The first two schemes, i.e. the modified-random-element-isodisplacement (MREI) and Cluster schemes, were developed at the emergence of such alloys in the mid-sixties by Chang and Mitra,[26] and by Verleur and Barker,[27] respectively. In the MREI scheme the bonds of a given species, i.e. AC- or BC-like, are all supposed to contribute to the same unique Raman line at a given alloy composition (1-bond→1-mode), irrespectively of their local environment. In contrast, the Cluster scheme distinguishes between like bonds depending on their 1ˢᵗ-neighbor environment at 3D, out of four possible clusters in a zincblende alloy (1-bond→4-mode). In each case, the equivalence between like elementary oscillators at a given alloy composition, i.e. the chemical bond itself in the MREI scheme and the chemical bond in its 1ˢᵗ-neighbor sphere in the Cluster scheme, is formalized by considering that the like oscillators in question are all immersed into the same uniform VCA-type continuum. This provides smooth composition dependence of the phonon frequencies, by construction. As for the intensity of a given Raman line, this simply scales as the fraction of corresponding elementary oscillators in the crystal.

The remaining scheme, i.e. the so-called Percolation scheme, as currently developed in our group in Metz,[28] deviates from the MREI and Cluster schemes in that it introduces a description of a random $A_{1-x}B_xC$ alloy in terms of a composite of the AC- and BC-like host regions, and not in terms of a uniform continuum. This naturally leads to distinguish between vibrations of like bonds depending on such two environments (1-bond→2-mode). Schematic comparisons between the contents of the Percolation scheme on the one hand and of the MREI and Cluster ones on the other hand are available, e.g. in Figs. 1 of Refs. **29** and **28**, respectively.

In principle the AC/BC-composite description implies singularities in the composition dependence of the Raman frequencies at each bond percolation threshold, at which critical alloy compositions the minor bonds connect into a pseudo-infinite treelike continuum.[30] In contrast, the AC/BC-composite description does not generate any singularity with respect to the Raman intensity aspect. More detail is given at a later stage. Persisting shortcomings in the current version of the Percolation scheme are concerned with the nature of the AC- and BC-like host regions, regarding both composition and length scale.

As far as SiGe is concerned, the 1-bond→1-mode MREI scheme is clearly undersized in view of the natural complexity of the Raman spectra. Besides, some of us have shown that the 1-bond→4-mode Cluster scheme is not devoid of conceptual ambiguities, and thus *in fine* misleading regarding the nature of the alloy disorder.[31] What remains is the Percolation scheme then.

The second objective of the present work is to see whether the Percolation scheme successfully used in our earlier work dedicated to $A_{1-x}B_xC$ zincblende semiconductor alloys,[28,29] may further apply to the diamond-type $Si_{1-x}Ge_x$ alloy.

In an attempt to extend the Percolation scheme to SiGe, we are aware that certain adaptation might be needed with respect to the original 1-bond→2-mode version, due to the zincblende→diamond change in the crystal structure. Moreover, somewhat paradoxically, the number of bond species enlarges from two in a ternary $A_{1-x}B_xC$ zincblende alloy (A-C and B-C) to three in the binary $Si_{1-x}Ge_x$ diamond alloy (Ge-Ge, Si-Ge and Si-Si). This is because all sites are likely to be occupied by the two atom species in $Si_{1-x}Ge_x$, while the C-sublattice remains unperturbed in $A_{1-x}B_xC$ zincblende alloys. Additional complication might arise in that the dispersion of the optical phonon is dramatically large in the pure Si and Ge crystals, i.e. of the order of several tens of cm$^{-1}$,[32] while it is almost negligible for the parents of all re-examined zincblende alloys so far, i.e. usually less than several cm$^{-1}$.[28] Last, we anticipate, in view of the unusually high complexity of the SiGe Raman spectra, that the Percolation scheme may not be adaptable to this alloy as long as the percolation-type environments of a bond remain undetermined at the microscopic scale.

The manuscript is organized as follows. In **Sec. II**, we introduce an improved version of the Percolation scheme for the reference $A_{1-x}B_xC$ zincblende alloys, in which the existing shortcomings regarding the exact nature of the AC- and BC-like environments of a bond are overcome. Based on this we design in **Sec. III-A** a seven-oscillator [1×(Ge-Ge), 4×(Si-Ge), 2×(Si-Si)] version of the Percolation scheme for random $Si_{1-x}Ge_x$ that naturally accounts for all intensity-interplays (**RI$_{1-2-3}$**). This considerably deviates from the presently prevalent six-oscillator [1×(Ge-Ge), 1×(Si-Ge), 4×(Si-Si)] picture reported in **Fig. 2b** (refer to the ovals). The Percolation-type re-assignment of the individual Raman lines realized in **Sec. III-A**, based on the composition dependence of their *intensities*, is independently secured in **Sec. III-B**, via an *ab initio* insight into the *frequencies* of bond-stretching modes along selected 3D-impurity motifs. These are taken as pseudo-linear, so as to remain in the spirit of the LCA, upon which the Percolation scheme relies. In **Sec. III-C** we give an overview of the SiGe Percolation-type Raman lineshapes over the whole composition domain, after *ab initio* calibration of the Ge-Ge, Si-Ge and Si-Si Raman efficiencies. Last, in **Sec. IV** we compare the SiGe and zincblende versions of the percolation scheme. A natural zincblende reference is GaAsP due to its similar lattice-mismatch as in SiGe. A striking difference between the two schemes is concerned with an inversion of the order of the like phonon modes in the Si-Si (**Sec. IV-A**) and Si-Ge (**Sec. IV-B**) multiplets with respect to the well-resolved Ga-P doublet. Such inversions are discussed via a comparative *ab initio* insight into the bond lengths and BZC-phonons of prototype impurity motifs that are directly transposable to the two crystal structures. Conclusions are summarized in **Sec. V**.

## II. Modified version of the Percolation scheme for the reference zincblende alloys

Our aim in this **Sec.** is to identify at the microscopic scale the AC- and BC-like percolation-type environments of a bond in the reference $A_{1-x}B_xC$ zincblende alloys, in terms of both composition and length scale. As we shall see, such forward step is needed before application of the Percolation scheme to SiGe. In working out such re-actualized version for zincblende alloys, which are polar (III-V, II-VI) in character, we focus on the non-polar transverse optical (TO) modes.[33] Indeed such purely mechanical oscillators assimilate in nature with the optical modes of the diamond-type SiGe alloy, basically a non-polar (IV-IV) one.[34] In fact, the required improvements will be concerned with the Raman intensity aspect only. Nevertheless, the Raman-frequency aspect is also described, for sake of completeness. Moreover this helps to understand a forecoming change in the terminology from the Percolation scheme to the 1D-Cluster scheme in the case of SiGe.

It is useful to recall that the Percolation scheme was originally developed to explain the disconcerting 1-bond→2-mode Raman behavior of $Zn_{1-x}Be_x$-chalcogenides.[35-37] Such behavior falls beyond the 1-bond→1-mode MREI scheme (undersized) and cannot be explained neither within the 1-bond→4-mode Cluster scheme (oversized) unless assuming a far from random Be↔Zn substitution,[28] which is totally unrealistic (Refs. [38-41]). In fact $Zn_{1-x}Be_x$-chalcogenides opened the attractive class of alloys involving the light first row elements from the periodic table in substitution, such as Be,[38-41] B,[42] N (Refs. [43, 44]) and O (Refs. [45-48]).[49] Such alloys are characterized by an unusually large contrast in their bond physical properties (see Fig. 1 in Ref. [49]). For example the bond length shortens by ~10% from Zn- to Be-based compounds, which goes with a doubling of the reduced shear modulus.[49] Altogether this generates a dramatic lattice distortion, as needed to accommodate the local strain due to the difference in bond lengths, leading to exacerbated vibrational properties. As such the novel class of alloys offers an opportunity to identify a sort of canonical Raman pattern for zincblende alloys in general. With this respect, $Zn_{1-x}Be_x$-chalcogenides are particularly interesting because they seem to be the only exceptions that can be grown as random alloys over the entire composition domain.[38-48]

In fact, we have shown, in a series of recent works, that the Percolation scheme worked out for $Zn_{1-x}Be_x$-chalcogenides successfully applies to all the leading systems (quoted within brackets hereafter) in the commonly admitted MREI/Cluster-based classification of the Raman and infrared spectra of zincblende alloys in four distinct types, i.e. two-mode (InGaAs), modified two-mode (InGaP), one-mode (ZnSeTe) and multi-mode (GaAsP).[28,29] This has lead to a unification of that classification. Coming to figures, $Zn_{1-x}Be_x$-chalcogenides exhibit a uniquely well-resolved 1-bond→2-mode Raman behavior, with a splitting of the order of ~40-50 cm$^{-1}$.[35-37] For the above-quoted less contrasted alloys, the splitting hardly exceeds ~10 cm$^{-1}$.[28,29]

### A. Raman intensity aspect

The intensity of a given Raman mode is directly monitored by its oscillator strength. In the current version of the Percolation scheme the available oscillator strength per bond, which scales as the corresponding bond fraction, divides between the two like modes of a given percolation doublet in proportion of the scattering volumes of the AC- and BC-like host regions. These are simply assumed to scale as the related bond fractions. For example, the intensities of the two like Raman features that form the AC-like (resp. BC-like) percolation doublet, referring to A-C (resp. B-C) vibrations in the AC-like and BC-like environments, do scale as $(1-x)^2$ [resp. $x \cdot (1-x)$] and $(1-x) \cdot x$ [resp. $x^2$], respectively. The two like modes from the same doublet thus have similar intensities at x~0.5, and the dominant mode at one end of the composition domain (x~0,1) turns minor at the other end. Such remarkable intensity-interplay (**RI**) was unambiguous observed with the Raman signal of the short Be-based bond of both ZnBeSe[35,36] and ZnBeTe[37] alloys.

In searching to identify the microstructure (composition, length scale) of the AC- and BC-like environments, we face a double constraint. First, regarding the length scale, it is remarkable that most of the phonon dispersion curves of a pure crystal, e.g. Ge or Si, can be reproduced by considering central and non-central inter-atomic force constants in between 1$^{st}$-neighbors only.[50] Accordingly, any assignment of the AC- and BC-like environments in the alloy beyond the 2$^{nd}$/3$^{rd}$ neighbors of a bond, might not be so realistic. Second, such environments should be defined at 1D, and not at 3D as in the Cluster model. This is to remain consistent with the LCA upon which the Percolation scheme relies.

In fact the above-mentioned scaling of the Raman intensities on the alloy composition is preserved simply by considering that the Percolation model distinguishes between the stretching of a bond depending on its AC- or BC-like 1$^{st}$-neighbor environment at 1D. For example $C(AC)B$ is identified as the 1D oscillator for the A-C bond-stretching in the BC-like environment. The corresponding fraction of oscillator then expresses as the probability of finding B ($x$) next to A ($1-x$) on the 1D (A,B)-like substituting sublattice, the two probabilities being independent in case of random A↔B substitution. The 1D oscillators for A-C bond-stretching in the AC-like environment, and for B-C bond-stretching in the AC- and BC-like environments are likewise identified as $C(AC)A$, $C(BC)A$ and $C(BC)B$, respectively.

### B. Raman frequency aspect

The description of an alloy in terms of a composite of the AC- and BC-like regions suffices to generate singularities in the Raman frequencies at the bond percolation thresholds (PT's), in principle. We recall that the bond PT's correspond to those critical alloy compositions at which the bonds forming the minor host region (AC- or BC-like) connect into a pseudo-infinite treelike continuum. Such self-connection is a pure statistical effect of the random A↔B site substitution.[30] It occurs at $x_B$~0.19 for the B-C bonds, and at the symmetrical value $x_A$~0.81 for the A-C ones in a random zincblende alloy (see Ref. **30**, chapter 1). For example, the two like branches from the A-C percolation doublet, as due to A-C vibration in the BC- and AC-like regions, are expected to exhibit a singularity at $x_B$ and $x_A$, respectively. Each singularity is due to a change in the topology of the minor host region from a dispersion of finite-size clusters into a treelike continuum, and more precisely, to the accompanying change in the internal structure from fractal-like, i.e. quasi-stable, to normal, i.e. smoothly x-dependent (see Ref. **30**, chapter 3), with concomitant impact on the phonon frequencies.

Technically, this was formalized as follows. We focus on e.g. the BC-like 1D-environment, for more clarity. A key result of the percolation site theory is that the BC-like treelike continuum that forms right at the bond PT ($x=x_B$) is a pure fractal, meaning that its internal structure is stable at all scales (see Ref. **30**, chapter 3). Another key result of this theory is that the internal structure of the BC-like continuum ($x>x_B$) modifies at the local scale as soon as departing from $x_B$, thus entering a so-called normal regime.[30] In the Percolation scheme we assimilate such BC-like continuum with a 'sub-mesoscopic' alloy within the main-macroscopic $A_{1-x}B_xC$ alloy, by using a renormalized MREI description. Accordingly the BC-like continuum takes a pseudo alloy composition y that is linearly-rescaled with respect to the actual alloy composition x, varying between 0 and 1 when x varies between the pure BC-like fractal ($x=x_B$), taken as a pseudo-parent (y=0), and the pure BC crystal (x=y=1).[35] A smooth x-variation generates in turn a smooth y-variation, with concomitant impact on the TO frequencies of the A-C and B-C bonds vibrating in the BC-like sub-alloy ($x_B \leq x \leq 1$). Besides, the percolation site theory indicates that the internal structure of the finite-size BC-like clusters that form the BC-like region below $x_B$ remains quasi-independent on the alloy composition x,[30] leading to quasi-invariance of the TO frequencies of the A-C and B-C bonds vibrating in such clusters ($0<x<x_B$). This is referred to as the fractal-like regime.

We must admit that such spectacular change in the 'TO-frequency vs. x' curve on each side of the bond PT, i.e. from quasi-stable (quasi-fractal regime) to smoothly x-dependent (normal regime), was unambiguously observed only with the highly contrasted ZnBeSe[35,36] and ZnBeTe[37] alloys so far. For less contrasted alloys, the percolation-type singularities in the Raman frequencies do not show up evidently.[28,29] Now, such singularities are the only justification for the terminology of a Percolation scheme. In this case, one may as well abandon the Raman frequency aspect,[51] and replace the terminology of a Percolation scheme by that of a 1D-Cluster scheme (by opposition with the 3D-Cluster scheme of Verleur and Barker – Ref. 28), in reference to the Raman intensity aspect only (see Sec. II-A).

### C. *Ab initio* protocol

Now, we must verify that the novel 1D assignment of the AC- and BC-like environments of a bond (see Sec. II-A) is technically consistent with our simple *ab initio* protocol in the dilute limits for the determination of the two input parameters needed per bond to implement the phenomenological Percolation scheme. These are the frequency of the impurity mode $\omega_{imp}$, plus the splitting $\Delta$ between the like TO modes of the same percolation doublet.[29]

We consider e.g. the BC-like percolation doublet. To access $\omega_{imp}$ we use a large AC-like supercell which contains a unique B impurity. This is the ultimate configuration that refers to an impurity (B) vibrating in the environment of the other species (AC-like). We search then for the vibration frequency of that impurity. There is only one triply-degenerate (triplet) mode for the isolated B atom. At 1D this naturally identifies with the *C(BC)A* oscillator. To access $\Delta$ we consider a pair of B impurities sitting nearby on the *fcc* substituting sublattice, forming a B–C–B pseudo-linear chain. This is the ultimate configuration that refers to an impurity (B) vibrating in its own environment (BC-like). The vibration pattern of the B–C–B pseudo-linear chain divides into two distinct series of BZC-like modes. The first series reduces to a singlet, corresponding to in-phase motion of the B atoms along the axis of the chain, against the intermediary C atom (q~0). Basically this refers to B-C bond-stretching along the pseudo-linear B-C-B chain, thus identified with *C(BC)B* at 1D. The second series consists of a multiplet including all possible bending modes of the B–C–B chain at q~0, in their in-plane and out-of-plane variants. Such bending modes of the BC-like chain do correspond to B-C bond-stretching perpendicular to the chain, namely inside the surrounding AC-like environment. As such they identify with the *C(BC)A* oscillator at 1D. In fact we have checked that the bending modes of the B–C–B chain do actually emerge close to $\omega_{imp.}$ (see e.g. Fig. 3 in Ref. 28). $\Delta$ is then estimated as the frequency gap between the impurity mode [*C(BC)A*] and the [*C(BC)B*] mode from the pair.

Incidentally, we did check that a pseudo-linear continuum of *percolating* B-C bonds, i.e. the natural *percolation*-type motif for B-C bonds vibrating in a BC-like environment, does provide a similar Raman pattern as the finite B-C-B chain mentioned above (compare the spectra in Figs. 9 and 3 of Refs. 52 and 53, respectively). In practice, the *ab initio* protocol is thus operated by using the latter, much more simple, motif.

We stress that the present re-actualized version of the Percolation scheme is not a predictive one, but instead offers a versatile framework for the interpretation of any 1-bond→multi-mode Raman pattern of any alloy. The 1D-oscillators can be expanded at will, in principle, depending on the number of observed Raman lines per bond and on the observed composition dependence of the Raman intensities. Such flexibility is crucial in our interpretation of the SiGe Raman spectra, as shown below.

### III. Percolation scheme for random SiGe

In our review of the SiGe Raman spectra, we did not find any hint of singularity in the composition dependence of the Raman frequencies at any bond PT.[54] This is not surprising, due to the rather small contrast in the bond physical properties of the Si and Ge crystals (see Sec. I). At the same time this means that the understanding of such curves falls beyond the scope of our simple Percolation scheme. Thus in our work we emphasize the Raman intensity aspect and adopt, from now on, the terminology of a 1D-Cluster scheme for SiGe (see Sec. II-B).

### A. 1D-Cluster assignment of the individual Raman lines: an intensity-based approach

In practice, we proceed as follows for the re-assignment of the SiGe Raman pattern at 1D.

First, we investigate roughly which Raman lines refer to which bond species. This is already known for the three main Raman features (refer to panels in Fig. 1), thus used as references. The assignment of the remaining four minor features is then inferred from the subtle inversions of dominance observed in between the main/reference Raman features (underscored in Figs. 1 and 3) and the minor/unknown ones (not underscored), in reference to $RI_{1-2-3}$. The basic idea is that the features involved in such interplays are coupled in some way, and hence are taken to involve the same bond-stretching. Accordingly, the low-frequency feature that counterbalances the main Si-Si one in $RI_1$ (at $x_{c1}$~0.70 – see Fig. 3c) and the two satellite ones involved by the main Si-Ge feature in **RI₂₋₃** (at $x_{c2}$~0.9 and $x_{c3}$~0.1 – see **Figs. 3a** and **3b**, respectively) would basically relate to Si-Si and Si-Ge bond-stretchings, respectively. In this case the Si-Si doublet and the Si-Ge triplet indicate that the corresponding bond-stretchings distinguish between two and three sorts of 1D-environments, respectively. Note that only three **RI**'s are identified, for four minor modes. One of the minor mode should thus remain unassigned, in fact the one corresponding to the dotted line in **Fig. 2**. Detail is given below.

The next issue is to identify the microstructure of the distinct 1D-environments per Si-Si and Si-Ge bond, in terms of both length scale and composition. We use two criteria: First (i) – simplicity, in view of the fact that BZC phonons essentially proceed from short-range interactions (see **Sec. II-A**), and second (ii) – an ability behind the corresponding fractions of elementary 'bond+environment' 1D-oscillators, as estimated on the realistic basis of a random Si↔Ge substitution, to reproduce the corresponding **RI's** at the observed alloy compositions, i.e. at $x_{c1}$ for the Si-Si doublet (**RI₁**), and at the symmetrical $x_{c2}$ and $x_{c3}$ values for the Si-Ge triplet (**RI₂, RI₃**).

First, we consider the Si-Si doublet. At the Si-dilute limit, most Si-Si bonds are isolated in the Ge-like matrix. Only a small fraction of these are sitting near other Si impurities. The trend is progressively reversed when increasing the Si content, until at a certain stage, the second environment becomes dominant.[30] Our view is that **RI₁** simply reflects such inversion of population. This naturally leads to a crude assignment of the low- and high-frequency components of the Si-Si doublet as being due to Si-Si bond stretching in Ge- and Si-like environments, respectively.

For a deeper insight, we investigate first whether it is possible to define such 1D-environments at the minimum length scale of 1st-neighbors, in reference to criterion (i). There are three possible Si-Si oscillators then, i.e. *Si(SiSi)Si*, *Si(SiSi)Ge* and *Ge(SiSi)Ge*, corresponding to Si-Si bond-stretching in pure-Si, (Si,Ge)-mixed and pure-Ge 1D-environments, respectively. The fractions of such 'bond+environment' 1D-oscillators simply express by weighting the Si-Si bond fraction (in reference to the atom species within the brackets), i.e. $(1-x)^2$, by the corresponding fractions of 1D-environments (in reference to the atom species outside the brackets), i.e. $(1-x)^2$, $2 \cdot x \cdot (1-x)$ and $x^2$, respectively, assuming a random Si↔Ge substitution. If we assume further that the Raman polarizability of the Si-Si bond is not dependent on its local neighborhood, in a first approximation,[55] then a quasi-perfect intensity-matching at $x_{c1}$ (cf. **RI₁**), in reference to criterion (ii), is achieved simply by assigning the low- and high-frequency Raman lines of the Si-Si doublet to the individual *Ge(SiSi)Ge* oscillator, and to the couple of [*Si(SiSi)Ge*, *Si(SiSi)Si*] oscillators taken as indiscernible, respectively. Perfect intensity-matching is then expected at 70 at.% Ge. This is consistent with experimental findings (refer to the Si-Si doublet at x=0.77 in **Fig. 1**), especially those of Rath *et al.* (refer to the morphological changes within the broad Si-Si band in Fig. 11 of Ref. **16**), and also with theoretical ones (in reference to the bottom panel of Fig. 5 in Ref. **21**). Accordingly the Si-Si bond-stretching distinguishes between full-Ge and alternative 1st-neighbor 1D-environments. The corresponding modes are labeled as $(Si-Si)_1^{Ge}$ and $(Si-Si)_1^{SiGe+Si}$ hereafter, respectively. In this compact notation, introduced already from **Fig. 1**, the main label specifies the nature of the bond-stretching, while the subscript and superscript refer to the length scale and to the composition of the 1D-environment, respectively.

We proceed in the same way for the Si-Ge triplet. We recall that, to our view, the striking similarity between the scenarios that develop for the main Si-Ge feature when entering the Si- and Ge-moderate/dilute limits implies assignments that are strictly similar in nature for the two side Si-Ge features involved in **RI₂₋₃**. In **Figs. 3a** and **3b** the two side features (not underscored) dominate the main one (underscored) in the Ge- and Si-dilute limits, where the environments of any Si-Ge bond are dominantly of the Si- and Ge-types, respectively. The corresponding side features are thus attributed to Si-Ge bond-stretching in Ge- and Si-like environments, respectively. This is consistent with our basic view that the side Si-Ge features and not the main one should connect to the Ge (x∼0) and Si (x∼1) LVM's (see **Sec. I**), in response to the long-standing enigma pointed out by Brya.[15]

Based on criterion (i), we consider first the possibility that the Si-Ge bond-stretching might distinguish between its three possible 1st-neighbor environments at 1D, as for the Si-Si bond-stretching. The 1D-oscillators associated with the [top, intermediate, bottom] Si-Ge branches, as mostly apparent at [moderate/dilute-Ge, intermediate composition, moderate/dilute-Si] would then be identified as [*Ge(GeSi)Ge*, *Si(GeSi)Ge*, *Si(GeSi)Si*], respectively. However, coming to criterion (ii), no chance exists, with the corresponding fractions of oscillators, to achieve intensity-matching at the observed $x_{c2}$ (**RI₂**) and $x_{c3}$ (**RI₃**) values. Intensity-matching would rather occur at 33 at.% of the minor species, and not around 10 at.%, in contradiction with experimental findings. We thus push the assignment to 2nd-neighbors, keeping in mind that the Ge- and Si-like 1D-environments should be identical in nature. In reference to criterion (ii), intensity-matching at $x_{c2}$ and $x_{c3}$ is then achieved simply by considering that the [top, intermediate, bottom] Si-Ge branches distinguish between [full-Si, all possible (Si,Ge)-mixed variants, full-Ge 1D] environments, respectively. The related fractions of oscillators express as $[2 \cdot x \cdot (1-x)^5, 2 \cdot x \cdot (1-x) \cdot \{[x+(1-x)]^4 - [x^4+(1-x)^4]\}, 2 \cdot (1-x) \cdot x^4]$ in this case, corresponding to intensity-matching between the central and peripheral features when entering the last 14 at.% of the composition domain. This is quasi ideally consistent with the data reported in **Figs. 2a** and **2b**, and also with the experimental findings of Rath *et al.* [corresponding to $x_{c2}$~0.86 and $x_{c3}$~0.14, as apparent in their detailed contour-modeling of the Si-Ge Raman signals using Lorentzian functions in Figs. 9(i) and 9(iii) of Ref. **16**, respectively]. The corresponding series of 1D-oscillators write as [*SiSi(GeSi)SiSi, GeSi(GeSi)SiSi + … + GeGe(GeSi)SiSi + … + GeGe(GeSi)GeGe, GeGe(GeSi)GeGe*], respectively, by using the extended notation, reducing to [$(Si-Ge)_2^{Si}$, $(Si-Ge)_2^{SiGe}$, $(Si-Ge)_2^{Ge}$] with the compact notation (see **Fig. 1**).

Now we come to the question of the remaining fourth, unassigned, minor oscillator (refer to the dotted line in **Fig. 2**). This relates to the top $(Si-Ge)_2^{Ge}$ branch. Soon after departing from the Ge-dilute/moderate limit, this branch seems to decompose into a sort of pseudo doublet. This is apparent in the experimental data shown in **Figs. 1** and **3c** (refer to the stars), and was detected theoretically by de Gironcoli already from x=0.28 on (see **Fig. 2a**). Within the scope of the 1D-Cluster scheme, such collapse reveals that the Si-Ge bonds vibrating in the Si-like environment turn sensitive to their 1D-environment at a length scale beyond 2$^{nd}$-neighbors. However, we must admit that we are unable to assign neither the actual microstructure nor the length scale of such 1D-environment. This is because the intensity-interplay between the two components of the pseudo-doublet is obscured by the side $(Si-Ge)_2^{SiGe}$ and $(Si-Si)_1^{Ge}$ features that become strong starting already from moderate Ge content. We simply note that the low-frequency component is stronger than its high-frequency counterpart at large Si content, and vice versa at large Ge-content (see **Fig. 1**). Such intensity-interplay basically indicates that the two components refer to Si- and Ge-like 1D-environments, respectively, being clear that such 1D-environments remain full Si-like up to 2$^{nd}$-neighbors, in reference to the native $(Si-Ge)_2^{Ge}$ mode.

The resulting 1D-Cluster scheme for SiGe is displayed in **Fig. 4**. The dotted line materializes the decomposition of the $(Si-Ge)_2^{SiGe}$ branch into the above-mentioned pseudo-doublet away from the Ge-dilute limit (refer to the oval on the left-hand side of **Fig. 4**). Otherwise, each individual branch is ideally represented by a straight plain line, the slope being defined so as to best adjust the experimental frequencies (see **Fig. 2a**). A significant deviation with respect to linearity is only observed for the central Si-Ge branch at moderate Ge content.

Altogether the SiGe 1D-Cluster scheme shown in **Fig. 4** provides a description of the SiGe Raman pattern in terms of seven modes [1×(Ge-Ge), 4×(Si-Ge), 2×(Si-Si)], as covered by six 1D-oscillators oscillators (refer to letters a, b, … f). This considerably deviates from the currently admitted six-oscillator [1×(Ge-Ge), 1×(Si-Ge), 4×(Si-Si)] picture at 3D reported in **Fig. 2b**, in which the $(Si-Ge)_2^{Ge}$ oscillator is omitted. In particular the 3D→1D re-assignment is apparent in the way the like oscillators are regrouped, as schematically indicated by ovals in the two figures. The gain behind such 3D→1D re-assignment is that the 1D-Cluster scheme is totally explicit regarding the intensity aspect, apart from the decomposition of the $(Si-Ge)_2^{Si}$ branch. This applies in particular to **RI$_{1-2-3}$** (see rectangles in **Fig. 4**). The intensity information is made apparent via an appropriate thickening of the individual branches in **Fig. 4**, in correspondence with the individual fractions of oscillators as specified on top of each branch. The corresponding information in **Fig. 2b** is restricted to the three main Ge-Ge, Si-Ge and Si-Si branches, and notably different with respect to **Fig. 4**, at least regarding the latter two branches.

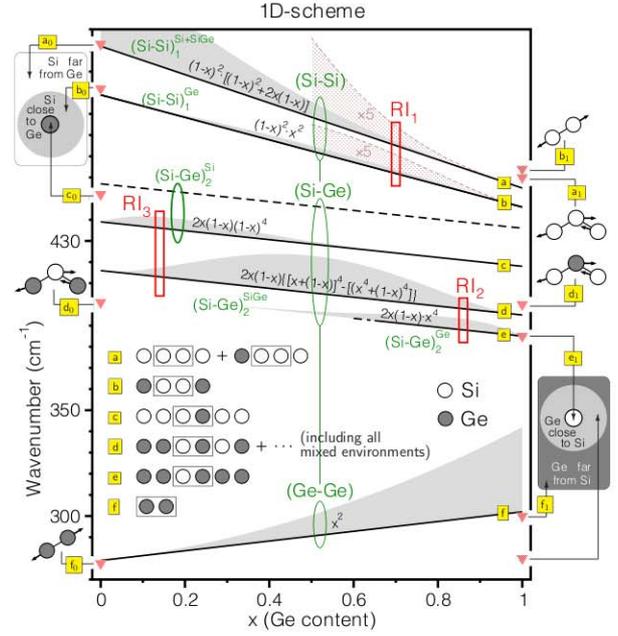

**Fig. 4**: **1D-Cluster scheme for SiGe**. This supports a seven-oscillator [1×(Ge-Ge), 4×(Si-Ge), 2×(Si-Si)] Raman pattern for Si$_{1-x}$Ge$_x$. The individual 1D-oscillators, identified as letters (a, b, …) are schematically represented in the body of the figure, using the same symbolic notation as in the panels of **Fig. 1**, for a direct comparison. The like phonon branches, referring to the same ultimate bond-stretching (as emphasized in the symbolic notation of each oscillator), are regrouped by ovals. Altogether, the resulting 1D assignment considerably deviates from the current 3D one as summarized in **Fig. 2b**, regarding both the nature and the number of oscillators. When available, the individual fractions of oscillators, which monitor the Raman intensities, are specified on top of each relevant branch, and visualized via an appropriate thickening of such branches (grey area), using the same scale as in **Fig. 2b**, for a direct comparison. The overall 1D-assignment, based on remarkable *intensity-interplays* (RI$_{1-2-3}$, see rectangles) at the theoretical critical Ge contents ($x_{c1}$=0.70, $x_{c2}$=0.86, $x_{c3}$=0.14) between like Raman lines, is independently validated by a direct *ab initio* insight into the *frequencies* of prototype vibration modes (identified by a letter, in reference to the considered oscillator, plus a subscript, in reference to the x value), as indicated on each side of the figure. In the two represented supercells that containing a unique impurity atom, a simple color code is used for Si (white) and Ge (dark grey), and to materialize the impurity-induced strain in the host medium (light grey). Note that the $(Si-Ge)_2^{Si}$ mode (refer to the oval on the left-hand side) decomposes into a distinct doublet away from the Ge-dilute/moderate limit (an overview is given in **Fig. 1**), which is taken into account by the adjunction of a dashed line.

### B. *Ab initio* insight into limit phonon-frequencies (x~0,1)

Independent insight into the 1D-Cluster re-assignment of the phonon branches is gained by *ab initio* phonon calculations using prototype impurity motifs taken as representative of the considered branches in their dilute limits (x~0,1), hosted by large (64-atom) supercells (red/clear triangles in **Figs. 2a** and **4**). The impurity motifs are taken as simple as possible, any impurity atom staying within the 2$^{nd}$-neighbor sphere of any other impurity atom from the same motif, each motif remaining beyond the 2$^{nd}$-neighbor sphere of the like motif from the next supercell. Altogether this insures that our supercells are sufficiently converged in size to mimic the phonon behavior of the motif as immersed into the infinite crystal.[56] When the considered impurity motif does not merely reduce to an isolated impurity, thus assimilating with a pseudo-linear chain, we focus on the bond-stretching mode along the chain (∥ chain), for sake of consistency with the LCA upon which the 1D-Cluster model relies (see **Sec. II**). Obviously such impurity motifs do not provide a ∥ chain mode only, but also ⊥ chain ones, with in-plane and out-of-plane variants. Nevertheless we have checked that the frequencies of the ⊥ chain modes related to a given LCA-type motif do roughly replicate, within less than 7 cm$^{-1}$, the frequency of the ∥ chain mode of another such motif. With this, no vibration mode of any motif is left unassigned.[57] The assignment of a given 1D-Cluster branch is eventually validated by a convergence of that branch onto the frequency of the ∥ chain mode of the related LCA-type motif. Such modes, as sketched out on each side of **Fig. 4**, are labeled by adding a subscript (0 or 1, in reference to the x value) to the letter (a, b …) representing each branch.

The phonon calculations are done after full relaxation of the supercells, i.e. of the lattice constant and of the atom positions, along the procedure detailed e.g. in Ref. **28**, using more specifically Eq. (2) therein. We use the density functional theory (DFT) and local density approximation (LDA) through the Ceperley-Alder[58] functional for the exchange-correlation energy in the SIESTA code (Ref. **59**). We take the separable Troullier-Martins[60] norm-conserving pseudopotentials with a basis set generalized to include double-zeta with polarization orbitals. The cutoff energy and *k*–grid cutoff are 360 Ry and 10 Å, respectively. When tested with the pure Si and Ge crystals, the obtained values of bond length, bulk modulus and BZC-phonon frequency with the SIESTA code compare fairly well with the experimental data and with earlier *ab initio* values given by the AIMPRO code, as shown in **Table I**. Detail regarding the AIMPRO code is given in the next subsection.

*Si-Si spectral range*. The 1-bond→2-mode Si-Si doublet requires *ab initio* insight into four limit frequencies. For the top $(Si-Si)_1^{SiGe+Si}$ branch, referring to Si-Si stretching in an undetermined SiGe- or Si-like environment, we consider two limit supercells, i.e. a pure-Si supercell, providing the bulk Si frequency [x~0, mode ($a_0$), 522 cm$^{-1}$], plus a Ge-like supercell with three quasi aligned Si impurities, searching then for the Si-Si stretching along the Si-Si-Si chain. This refers to Si-Si stretching in a (Si,Ge)-mixed LCA-type 1$^{st}$-neighbor (1) environment [x~1, mode ($a_1$), 460 cm$^{-1}$]. For the bottom $(Si-Si)_1^{Ge}$ branch due to Si-Si stretching in a full-Ge (1)-environment, we consider a Ge-like supercell with an isolated Si-Si bond, searching for its stretching mode [x~1, mode ($b_1$), 464 cm$^{-1}$], plus a Si-like supercell with an isolated Ge atom being interested then in the

|  | Nearest Neighbor Distance (Å) | | Bulk Modulus B (GPa) | | $\omega_0$ (cm$^{-1}$) | |
| --- | --- | --- | --- | --- | --- | --- |
|  | *Ab initio* | Exp. | *Ab initio* | Exp. | *Ab initio* | Exp. |
| Si | 2.089 *2.335*[a] | 2.327[b] | 100.4 | 99.9[c] 99.7[b] | 522 *518.1*[a] | 520.2[e] |
| Ge | 2.183 *2.417*[a] | 2.443[b] | 70.1 *75.2*[a] | 78.1[d] | 304 *300.9*[a] | 300.7[e] |

[a]Ref. **61**, [b]Ref. **62**, [c]Ref. **63**, [d]Ref. **64**, [e]Ref. **65**

**Table I:** *Ab initio* bond lengths, bulk moduli and optical phonon frequencies of the pure Si and Ge crystals presently obtained with the SIESTA code, as compared with experimental values and with earlier *ab initio* values obtained with the AIMPRO code (cf. italics).[61]

Si-Si stretching close to that impurity [x~0, mode ($b_0$), 502 cm$^{-1}$].

In the Si-dilute limit (x~1), the two Si-Si stretching modes ($a_1$)–($b_1$) converge to 462 cm$^{-1}$ within less than 2 cm$^{-1}$, which is globally consistent with the trend observed in **Fig. 2a**. However, in contrast with earlier calculations of Franz *et al*. (see **Sec. V-C** of Ref. **17**), we are not able to resolve the experimental splitting of ~9 cm$^{-1}$ between the two branches. For a more decisive insight we turn to the Ge-dilute limit (x~0) where the splitting is expected to be much larger, i.e. of ~30 cm$^{-1}$, based on the experimental observations of Brya.[15] As a matter of fact our *ab initio* calculations do reveal a distinct Si-Si stretching mode close to the isolated Ge [($b_0$), the vibration pattern is shown in **Fig. 6b**]. This emerges at ~20 cm$^{-1}$ below the bulk-like Si-Si stretching far from Ge ($a_0$), in reasonable agreement with experimental findings (see **Fig. 2a**). Altogether, such *ab initio* insight secures the 1D-Cluster assignment of the Si-Si doublet. Incidentally, the existence of a Ge-related mode at ~500 cm$^{-1}$, in reference to ($b_0$), was predicted by Maradudin[19] (see **Fig. 2a**). However, it was unsuspected that such mode is not due to a vibration of the Ge impurity itself, but of the nearby Si-Si bonds.

*Si-Ge spectral range*. In principle, an insight into six end frequencies is required to test the basic 1-bond→3-mode Si-Ge pattern, if we omit the decomposition of the $(Si-Ge)_2^{Si}$ branch. However, our limitation to simple impurity motifs, so as to fulfill a basic condition regarding the convergence of our *ab initio* calculations (see above), excludes the analysis of those Si-Ge branches referring to 1D-environments formed with impurities only, such as $(Si-Ge)_2^{Si}$ at x~1 and $(Si-Ge)_2^{Ge}$ at x~0. The remaining four limit frequencies are accessed by using two symmetrical pairs of impurity motifs. An isolated Si atom in Ge gives access to Si-Ge stretching in a full-Ge environment [x~1, mode ($e_1$), 385 cm$^{-1}$], corresponding to the LVM of Si in Ge. Its counterpart on the Si side, i.e. the Ge-LVM, is likewise accessed by considering an isolated Ge atom in Si, focusing on the actual impurity vibration [x~0, mode ($c_0$), 451 cm$^{-1}$]. The limit frequencies of the intermediate Si-Ge branch, due to Si-Ge stretching in a (Si,Ge)-mixed environment, are accessed by using a pair of impurities in 2$^{nd}$-neighbor positions, searching for the in-pair impurity-stretching against the intermediate host atom [refer to the ($d_0$) and ($d_1$) modes, at 401 and 400 cm$^{-1}$, respectively].

Altogether, each branch benefits from at least one *ab initio* insight, and the related frequencies match remarkably well the three Si-Ge branches. This secures the Si-Ge 1D-Cluster pattern. The ($d_1$)–($e_1$) splitting around ~400 cm$^{-1}$ has already been identified by Franz *et al.*,[17] but using a different motif for ($d_1$) (see Ref. 50). In contrast the ($c_0$)–($d_0$) counterpart remained unexplored so far. As we did anticipate (see Sec. III-A), the Ge-LVM ($c_0$) and Si-LVM ($e_1$) are connected with the top and bottom Si-Ge branches, respectively, and not with the central/main one. In particular, at x~0, where the two Si-Ge branches are well separated, the central/main Si-Ge branch unambiguously extrapolates to ($d_0$) (see Fig. 2a). This brings decisive insight into the nature of this branch, solving at the same time the long-standing enigma raised up by Brya.[15]

$(Si-Ge)_2^{Si}$ *fine structure*. The collapse of the $(Si-Ge)_2^{SiGe}$ branch into a pseudo-doublet away from x~0 cannot be checked directly via our present *ab initio* calculations, as discussed above. For a direct insight we refer to the dominant vibration patterns sketched out in the body of Fig. 2b, as identified by Grein and Cardona.[20] In each case the bond-stretching is emphasized by an oval, for more clarity.

As expected, the extreme (1, $2_1$) and ($4_2$, $5_1$) pairs of vibration patterns, at low and high frequency, respectively, do basically refer to Si-Ge and Si-Si stretching, respectively. Moreover, in each case the phonon frequency increases when the local environment becomes more Si-like. This is consistent with our generic 1D-Cluster assignment of such doublets in terms of $[(Si-Ge)_2^{Ge}, (Si-Ge)_2^{SiGe}]$ and $[(Si-Si)_1^{Ge}, (Si-Si)_1^{SiGe+Si}]$, respectively. The key question then is whether the features in-between, corresponding to the vibration patterns ($3_1$, $4_1$), do basically refer to Si-Ge stretching, as we presume, or to Si-Si stretching? Clearly both patterns refer to Si-Ge stretching (see the ovals) in a Si-like environment (pay attention to the local composition). This conforms to our view that such modes basically proceed from the original $(Si-Ge)_2^{Si}$ branch.

*Ge-Ge spectral range*. The two limit frequencies of the apparently unique Ge-Ge branch are accessed by considering a pure-Ge supercell [x~1, mode ($f_1$), 304 cm$^{-1}$], plus a Si-like supercell containing a pair of 1$^{st}$-neighbor Ge impurities, searching then for the Ge-Ge stretching [x~0, mode ($f_0$), 281 cm$^{-1}$]. The as-obtained *ab initio* values match the experimental ones within less than 5 cm$^{-1}$.

By curiosity, we have run further *ab initio* calculations in search of a possible extra Ge-Ge mode. In doing so we considered a Ge-like supercell containing a unique Si impurity, in reference to the symmetrical supercell successfully used to evidence the ($a_0$)–($b_0$) splitting for Si-Si. In fact the Si-induced tensile strain generates a local Ge-Ge stretching nearby Si at significantly lower frequency [x~1, mode at ~280 cm$^{-1}$] than the bulk-like Ge-Ge stretching far off Si [x~1, mode ($f_1$), 304 cm$^{-1}$], as schematically indicated in Fig. 4. The vibration pattern of such Ge-Ge stretching close to Si is just identical to that found for the local Si-Si mode ($b_0$) close to an isolated Ge atom in Si, as sketched out in Fig. 6b. We deduce that the mono-mode Ge-Ge behavior apparent in the Raman spectra may not be intrinsic but due to a screening of an actual multi-mode pattern similar to the Si-Si one. The reason why the Ge-Ge Raman signal does not develop into a proper multi-mode pattern is not clear yet. We simply note along with Rath *et al.*[16] that the Ge-Ge spectral range is quasi-constantly blurred by a parasitical disorder-induced phonon density of states. This materializes into a distinct quasi-amorphous (q-a, refer to Fig. 1) band covering a broad frequency range (120 – 310 cm$^{-1}$) already from low Si content (see Fig. 2 of Ref. 16).

In brief, our simple *frequency-based ab initio* protocol in the dilute limits secures the *intensity-based* 1D-Cluster assignment of the phonon branches (see Sec. III-A) reported in Fig. 4. Further it reveals a predisposition of the Ge-Ge bond to exhibit a multi-mode Raman pattern, like Si-Si. However, this seems to be hampered by parasitical disorder-induced effects.

### C. Contour modeling of the Raman lineshapes within the 1D-Cluster scheme

An overview of the 1D-Cluster SiGe Raman lineshapes is derived by taking the imaginary part of a sum of six Lorentzian functions representing the six basic [1×(Ge-Ge), 3×(Si-Ge), 2×(Si-Si)] harmonic oscillators, using the frequency (plain lines) / intensity (in reference to the fractions of oscillators as specified on top of each branch) information available per oscillator in Fig. 4. We neglect the decomposition of the $(Si-Ge)_2^{Si}$ branch away from the Ge-dilute/moderate limit, where anyway the contribution of this mode to the overall Raman signal is negligible.

We emphasize that the fractions of oscillators merely act as weighting factors applying to the intrinsic Si-Si, Si-Ge and Ge-Ge Raman efficiencies. Now, these remain to be determined. An experimental probing is problematic already because the perfect zincblende SiGe crystal, formed with Si-Ge bonds only, does not exist in reality. We thus turn to a direct *ab initio* calculation of the Raman spectra of the diamond-type Si and Ge crystals and of the zincblende-type SiGe compound. The calculations are done by using a pseudopotential spin density-functional supercell code (AIMPRO),[66] along the local exchange-correlation parametrization by Perdew and Wang,[67] taking the potentials for Si and Ge as proposed by Hartwigsen *et al.* (Ref. 68). In doing so, we follow the exact procedure as earlier optimized for SiGe in Ref. 61. The resulting Si, Ge and SiGe *ab initio* Raman spectra are reported in the inset of Fig. 5. Supercells of the same size were used for a direct comparison of the intrinsic (Ge-Ge, Si-Ge, Si-Si) Raman efficiencies. These scale as the corresponding Raman intensities, i.e. approximately as (1, 2/3, 1/2).

The only adjustable parameter in the calculation of the Raman spectra of the disordered SiGe alloy is the phonon damping per mode. The same value is taken for the like Raman lines of a given 1D-Cluster multiplet, simply because such lines ultimately refer to the same bond-stretching. Experimental support arises in that the like Si-Ge (see **Figs. 3a** and **3b**) and Si-Si (see **Fig. 3c**) Raman peaks do exhibit similar linewidths when they coexist with similar intensities. Further, we consider that the individual Ge-Ge, Si-Ge and Si-Si dampings are not composition-dependent. This is not true in particular for the Ge-Ge mode, in view of the strong composition dependence of both its linewidth and asymmetry, as carefully measured by Rath *et al*. (see Fig. 5b of Ref. **16**). Technically, such dependence can be easily incorporated in the calculation of the Raman spectra, i.e. via an appropriate composition-dependent asymmetrical damping.[69] However, in this case, a similar dependence should be also considered for the Si-Si and Si-Ge modes then. Now, we are not aware that such data exist in the literature. Thus, considering it better to treat the three modes on equal footing in our basic calculation of the SiGe Raman spectra, merely an indicative one, we stick to the crude approximation of stable dampings.

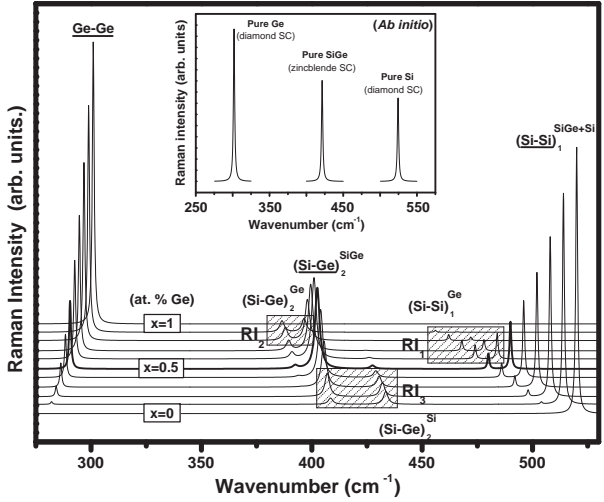

**Fig. 5**: **1D-Cluster SiGe Raman lineshapes**. Reference 1D-Cluster Raman lineshapes for random $Si_{1-x}Ge_x$ as derived by taking the imaginary part of a classical six-oscillator [1×(Ge-Ge), 3×(Si-Ge), 2×(Si-Si)] dielectric function, using the phonon-frequencies (plain lines) and fractions of oscillators (as specified on top of each branch) available in **Fig. 4**. The decomposition of the $(Si-Ge)_2^{Si}$ mode is neglected in a first approximation. The fractions of oscillators, which monitor the Raman intensities, merely act as weighting factors applying to the Ge-Ge, Si-Ge or Si-Si intrinsic Raman efficiencies. These are determined from *ab initio* calculation of the Raman spectra of pure Ge (diamond), Si (diamond) and SiGe (zincblende) supercells (SC's) of the same size. The corresponding curves are shown in the inset. Eventually, an experimental observation that the main Ge-Ge, Si-Ge and Si-Si Raman lines exhibit comparable intensities at x∼0.5 (see **Fig. 1**) is well-reproduced by taking phonon dampings that scale as 1:2:1. A small phonon damping (of 1cm$^{-1}$ for the reference Ge-Ge line) is used, for a clear overview of the remarkable intensity-interplays **RI$_{1-3}$** in between like Raman lines, as emphasized by hatched areas (refer to **Fig. 3** for a comparison with experiment).

We estimate the Si-Si ($\gamma_{Si}$), Si-Ge ($\gamma_{SiGe}$) and Ge-Ge ($\gamma_{Ge}$) dampings from experiment at x∼0.5. The three bond species coexist in similar proportions in the crystal, and thus exhibit comparable Raman signals. This is apparent in **Fig. 1**, in the spectra of Rath *et al*. (see Fig. 1 of Ref. **16**), and also in the theoretical Raman spectra of de Gironcoli (see Fig. 5 of Ref. **21**).[70] Such trend is well reproduced in our simulations by taking $\gamma_{SiGe} \sim 2\gamma_{Si} \sim 2\gamma_{Ge}$.

A representative set of as-obtained 1D-Cluster Raman lineshapes for random $Si_{1-x}Ge_x$ is displayed in **Fig. 5**. The reference curve at x∼0.5 is emphasized. A small damping is taken ($\gamma_{Si} \sim 1 cm^{-1}$) for optimal resolution of neighboring features. The **RI**'s at **$x_{c1}$**∼0.70 (**RI$_1$**), **$x_{c2}$**∼0.90 (**RI$_2$**) and **$x_{c3}$**∼0.10 (**RI$_3$**) are naturally accounted for (refer to the shaded areas).

Note that our various approximations have no impact on such interplays. They only modify the relative strengths of Raman lines due to distinct bond-stretching, while the **RI**'s take place in between like Raman lines referring to the same bond-stretching. When approaching the moderate/dilute limits, the marked composition dependence of the Ge-Ge phonon damping might introduce some distortion with respect to the theoretical intensity ratios between the Ge-Ge and (Si-Ge, Si-Si) Raman lines. However, this will just emphasize basic trends that the Ge-Ge signal becomes either hardly detectable (x∼0) or largely dominant (x∼1).

Summarizing, in spite of its simplicity, our 1D-Cluster scheme provides fair understanding, on a quantitative basis, of the natural complexity of the $Si_{1-x}Ge_x$ Raman spectra in their composition dependence, apart from the $(Si-Ge)_2^{Si}$ decomposition.

## IV. SiGe (diamond) vs. GaAsP (zincblende) versions of the Percolation scheme

An interesting question, then, is how the present version of the 1D-Cluster scheme for diamond SiGe does compare with the current 1-bond→2-mode version of a zincblende alloy? A natural zincblende reference with this respect is GaAsP, owing to a similar lattice mismatch as SiGe, i.e. ∼3-4%. The local relaxations are thus comparable, as the microscopic strain. Precisely, it appears in retrospect that the microscopic strain is the one crucial ingredient which governs the ordering of the like phonon sub-branches in the 1D-Cluster doublets of all re-examined zincblende alloys so far. Detail is given below. Additional interest arises in that SiGe and GaAsP are highly contrasted regarding the dispersion of their optical modes. These are nearly dispersionless in GaAs (∼15 cm$^{-1}$, Ref. **71**) and GaP (∼1.5 cm$^{-1}$, Ref. **72**), and strongly dispersive in Si (∼60 cm$^{-1}$, Ref. **32**) and Ge (∼30 cm$^{-1}$, Ref. **32**). This offers a unique opportunity to appreciate to which extent the dispersion effect might challenge the zincblende version of the 1D-Cluster scheme, based on the microscopic strain effect.

A basic difference between the SiGe (see **Fig. 4**) and GaAsP (see Fig. 7 of Ref. **29**) 1D-Cluster schemes is the opposite order of the like phonon branches in each multiplet. If we consider the distinct Ga-P doublet of GaAsP, the low (high) frequency branch refers to the host region mainly formed with the short (long) bond, i.e. GaP-like (GaAs-like). In contrast, the like phonon branches in each of the Si-Ge and Si-Si multiplets are ranked from bottom to top in the sense of more Si-rich host environments, as mainly formed with short Si-based bonds. Such Si-Si and Si-Ge inversions are successively discussed hereafter.

### A. Inversion of the Si-Si branches – A dispersion effect discussed at the Si-parent limit (x~0)

We tackle the issue of the Si-Si inversion with respect to Ga-P at the Si/GaP-parent limit ($x$~0), where the Si-Si doublet is best resolved, via *ab initio* calculations. The ultimate supercell providing simultaneous access to the two modes of a given doublet is a Si/GaP-like one containing a single Ge/As-impurity, as sketched out in Fig. 6a. Our aim is to compare the lattice relaxation/dynamics of the host medium close to the impurity and far from it in the two systems, after full relaxation of the supercells. The medium-related bond length distributions and BZC-phonon curves of the (Ge,As)-doped (Si,GaP)-like supercells are shown in Figs. 6a and 6b, respectively. Corresponding features in the two panels are identified by using the same simplified labeling with no subscript, for a direct link.

The $(Si-Si)^{Ge}/(Ga-P)^{As}$ host bonds near the Ge/As impurity (refer to the superscript) suffer a local compression due to the long (Ge-Si, Ga-As) impurity bonds (refer to the grey area around the isolated impurity in Fig. 6a). This gives rise to a distinct shoulder on the short-bond side of the main $(Si-Si)^{Si}/(Ga-P)^{P}$ bulk-like feature in Fig. 6a. Transposed to the lattice dynamics (Fig. 6b), one would expect that the host bonds close to the impurity do generate a distinct localized phonon at a higher frequency than the BZC-like ($q$~0) mode from the bulk. Indeed, an intuitive rule at 1D is that *the force constant of a bond reinforces when the bond is shortened, leading to hardening (blue-shift) of the phonon mode, and vice-versa*. This is actually so for GaAsP, the blue-shift being of the order of 10 cm$^{-1}$, but not for SiGe, a red-shift of ~25 cm$^{-1}$ being detected instead. Such apparent anomaly for SiGe can be explained only if the compression-induced hardening effect (blue-shift) is compensated by a larger confinement-induced softening effect (red-shift).

As extensively discussed by Rücker and Methfessel (Ref. 3), the central idea behind the confinement effect is that the host medium constitutes an obstacle to the propagation of a 'foreign' (impurity) mode, simply because it does not naturally vibrate at the same frequency. Therefore the impurity mode remains confined onto the impurity motif. As such, it cannot be described in terms of a nominal $q$~0 Raman-allowed mode, corresponding to a quasi-infinite correlation length. It is currently decomposed into a series of elementary plane waves involving disorder-induced theoretically forbidden $q \neq 0$ modes. A dominant $q$~0 (BZC) character is nevertheless presumed, corresponding to out-of-phase vibration (optical character) of the two intercalated *fcc* sublattices taken as quasi-rigid (BZC character). Such picture has been formalized into the well-known spatial correlation model (SCM),[73,74] in which the series runs over all possible $q$ values from the BZC to the

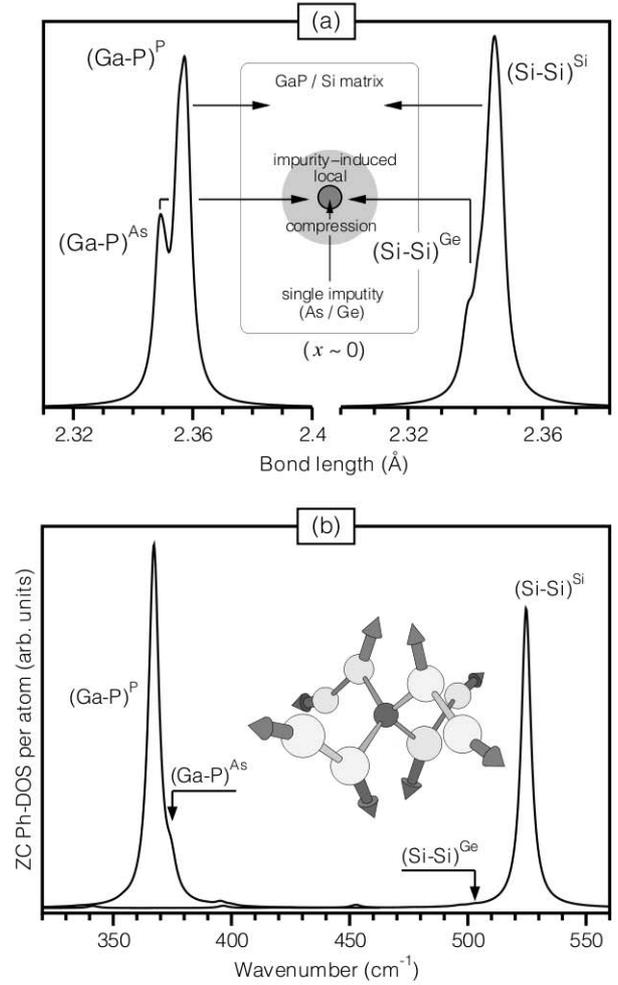

**Fig. 6**: **SiGe (diamond) vs. GaAsP (zincblende) comparison – Inversion of the Si-Si doublet**: *Ab initio* bond lengths distributions [panel (a)] and phonon density of states at the Brillouin zone-center [ZC Ph-DOS, panel (b)] referring to the host Ga-P/Si-Si bonds of similar GaP/Si-like (left/right) supercells containing a unique As/Ge impurity (x~0), as sketched out in panel (a). The isolated As/Ge impurities produce a local compression of the host medium (refer to the grey area). The *ab initio* vibration pattern of the corresponding localized matrix-like $(Ga-P)^{As}/(Si-Si)^{Ge}$ modes [in reference to ($b_0$) in Fig. 4 for SiGe] is schematically represented in panel (b). The host atoms beyond the 2$^{nd}$-neighbors of the impurity remain motionless (not shown). Corresponding bond length/phonon features in panels (a) and (b) are labeled by using the same simplified 1D-Cluster terminology with no subscript, for a direct link.

Brillouin zone edge (BZE), with a decreasing weight of the elementary modes when $q$ gets closer to the BZE. In most semiconductors, the optical phonons exhibit a negative dispersion, so that the confinement effect usually leads to a softening (red-shift) with respect to the nominal BZC mode of a pure crystal.

Now, such a drastic red-shift as ~25 cm$^{-1}$ cannot be explained within the SCM, i.e. by assuming that the impurity mode retains a dominant BZC-like character. Indeed, even for a confinement of the highly dispersive Si-Si optical mode at the ultimate scale of the lattice constant, the red-shift predicted by the SCM does not exceed ~5 cm$^{-1}$ (see Fig. 1 of Ref. 73).

The only way to take full advantage of the phonon dispersion is to suppose that the localized Si-Si impurity mode behaves more like a BZE mode than like a BZC one. Such BZE mode can be identified via its vibration pattern, considering that not only the two $fcc$ sublattice should vibrate out of phase (optical character), but also the 1$^{st}$-neighbors on each $fcc$ sublattice (BZE character). For a direct insight we have sketched out in Fig. 6b the *ab initio* vibration pattern of the Si-Si stretching mode close to Ge. We have checked that the Ga-P vibration pattern close to As is similar. The atom displacements are quasi null beyond the 2$^{nd}$-neighbors of the isolated impurity (not shown), which establishes the local character of such impurity-induced modes. Further we observe that the Si (P) atoms immediately connected to the isolated Ge (As) impurity, thus located on the same $fcc$ sublattice, exhibit anti-phase displacements two-to-two. The same holds true for the Si/P atoms sitting in 2$^{nd}$-neighbor positions on the same Ge/As-like $fcc$ sublattice. Altogether this testifies for an effective BZE-like character of the Si-Si and Ga-P local modes close to the Ge/As impurities, as expected.

The dispersion-induced phonon-softening is maximum for BZE phonons, i.e. of the same order as the magnitude of the BZC-BZE dispersion. In the case of GaAsP the hardening effect due to the impurity-induced compressive strain is not challenged for all that, due to the absence of Ga-P dispersion. In contrast the dispersion-induced softening should easily screen the strain-induced hardening in SiGe, a rather moderate one in fact if we refer to GaAsP (of ~10 cm$^{-1}$), owing to the large Si dispersion (~60 cm$^{-1}$). In this case the Si-Si local mode near Ge (BZE-like) falls over the Si-Si parent-like one far off Ge (BZC-like) at x~0. This, we believe, is the origin of the Si-Si inversion with respect to Ga-P.

### B. Inversion of the Si-Ge branches – A lattice relaxation effect discussed at the Si-dilute limit (x~1)

We discuss the Si-Ge inversion with respect to Ga-P at a dilute limit. All Si-Ge modes are of the impurity type then, thus suffering a similar dispersion effect. As such, the dispersion effect is virtually excluded from the discussion. Only the microscopic strain needs to be considered. Everything comes down to a comparison of the lattice relaxations in the zincblende (GaAsP) and diamond (SiGe) lattices then, from the point of view of the dilute (Si-Ge)/(Ga-P) bonds. The lattice dynamics immediately follows in principle, using the basic rule quoted in italics above. Now, at the (Ga-P)–dilute limit, the short Ga-P bonds are dispersed into the GaAs-like host medium with a large lattice constant. The equivalent situation for the Si-Ge bond is achieved at the Si-dilute limit only, the minor Si-Ge bonds being shorter than the host Ge-Ge ones. The Si-Ge vs. Ga-P confrontation is thus placed at this limit.

For a direct Si-Ge vs. Ga-P comparison we use impurity motifs that are transposable from the diamond structure to the zincblende one. An inevitable drawback is that such comparison is limited to the bottom and intermediate Si-Ge branches only. The top Si-Ge branch cannot be addressed since the related impurity motif, i.e. a pseudo-linear chain of five Si atoms with an intermediate Ge atom (see Fig. 4), does not transpose to the zincblende structure. In practice, we select those two motifs currently used to run the zincblende version of our *ab initio* protocol (see **Sec. II**). These consist of one isolated (Si/P) impurity (1-imp.) plus one pair of 2$^{nd}$-neighbor impurities (2-imp.), to be immersed in Ge/GaAs-like supercells. The impurity-related distribution of bond lengths and BZC-phonon curves after full relaxation of the supercells are shown in **Figs 7a** and **7b**, respectively. Again, corresponding features in the two data sets are identified by using the same simplified labeling with no subscript, for a direct link. As expected, there is a striking difference between the zincblende and diamond lattice relaxations, with concomitant impact on the lattice dynamics.

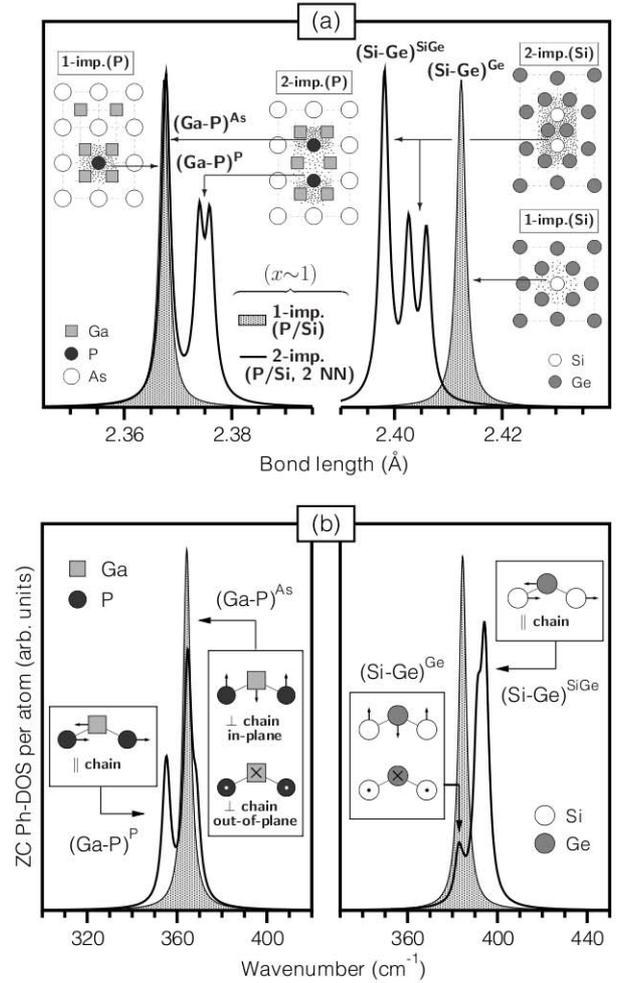

**Fig. 7**: SiGe (diamond) vs. GaAsP (zincblende) comparison – **Inversion of the Si-Ge multiplet**: *Ab initio* bond lengths distributions [panel (a)] and phonon density of states at the Brillouin zone-center [ZC Ph-DOS, panel (b)] referring to the impurity bonds of similar GaAs/Ge-like (left/right) supercells (x~1) containing either an isolated P/Si impurity (grey curves) or a pair of such impurities sitting in 2$^{nd}$-neighbor positions (clear curves), as sketched out in panel (a). Clouds of dense/dispersed dots in such schemes refer to longer/shorter impurity bonds. Note that the situations are opposite for the GaAsP and SiGe schemes. Corresponding bond length/phonon features – or set of features – in panels (a) and (b) are labeled by using the same simplified 1D-Cluster type terminology with no subscript, for a direct link. In panel (b) only schematic vibration patterns are displayed [in reference to ($d_1 - e_1$) in Fig. 4 for SiGe], for a direct SiGe vs GaAsP comparison (qualitative). A standard notation is used for backward (⊠ or ⊗) and upward (⊙) atomic motions.

The short Si-Ge bonds in Ge are shorter (dark area) in the Si-rich (2-imp.) region, and longer (clear area) in the Si-poor (1-imp.) one (**Fig. 7a/right**), thus vibrating at higher and lower frequency (recall the basic rule in italics), respectively (**Fig. 7b/right**). This conforms to intuition. Indeed the connection of the Si-impurity motifs to the surrounding Ge-like matrix involves eight short Si-Ge bonds for 2-imp. against four only for 1-imp. The Si-rich domain (2-imp.) thus offers a stronger resistance to the medium-induced tensile strain, and as such retains more efficiently the naturally short Si-Ge bond length.

Surprisingly, the trend is opposite for Ga-P, the short Ga-P bonds being longer (dark area) inside the P-Ga-P (2-imp.) chain and shorter (clear area) at its extremities, or around an isolated-P (1-imp.) atom (**Fig. 7a/left**), with concomitant impact on the Ga-P phonon frequencies (**Fig. 7b/left**). In fact, the latter two series of Ga-P bonds exhibit similar bond length, indicating that the local relaxation *outside* a given P-impurity motif does not depend on the arrangement of the P atoms *inside* that motif. This is consistent with a current observation by extended x-ray absorption fine structure measurements in zincblende alloys that the substituting $fcc$ sublattice remains nearly undistorted.[75-78] The local strain is merely accommodated by a distortion of the invariant $fcc$ sublattice. This was formalized into a model by Balzarotti *et al*. (Ref. **79**). As the P atoms stay at the nodes of the undistorted $fcc$ GaAs-like sublattice, with a large lattice constant, the local tensile strain inside the P–Ga–P (2-imp.) motif can only be accommodated by a mere in-plane motion of the central Ga atom towards the P-pair. Such relaxation is not as efficient as that achieved at the two extremities of the P–Ga–P chain or around the isolated-P atom (1-imp.), where three and four unconstrained Ga atoms are available per P atom, respectively. Accordingly, the former in-chain Ga-P bonds remain longer than the latter ones.

In summary, we attribute the opposite order of the Si-Ge and Ga-P branches to a difference in nature between the local lattice relaxations in the diamond and zincblende alloys. In the first case, all sites are equivalent in the relaxation process, so that an impurity motif tends to shrink or expand as a whole in order to retain the native bond lengths of its constituting species. In contrast, the zincblende relaxation process is constrained to a strict condition that the substituting $fcc$ sublattice should remain undistorted. This suffices to generate a counter-intuitive trend in the lattice relaxation that short bonds tend to be longer in their own environment than in the environment of the other (long) species. This reflects in the order of the two sub-branches of a 1D-Cluster doublet. This, we believe, is the origin of the Si-Ge inversion with respect to Ga-P.

### IV. Conclusion

A re-actualized version of the phenomenological, i.e. LCA-based, Percolation scheme, which has recently led to a unification of the classification of the Raman spectra of the III-V and II-VI $A_{1-x}B_xC$ zincblende semiconductor alloys,[28,29] is applied to random $Si_{1-x}Ge_x$, the leading group-IV semiconductor alloy with diamond structure. In this so-called 1D-Cluster version, the current shortcomings regarding the assignment of the individual Raman lines, which relate to the composition and length scale of the 1D-environment of a bond, are overcome.

Remarkable *intensity*-interplays within both the Si-Si ($RI_1$) and Si-Ge ($RI_{2-3}$) spectral ranges when reaching the Si- and Ge-dilute/moderate limits, known from the literature, are used to reassign the individual SiGe Raman lines into a proper seven-oscillator [1×(Ge-Ge), 4×(Si-Ge), 2×(Si-Si)] version of the 1D-Cluster scheme. This strongly deviates from the currently admitted six-oscillator [1×(Ge-Ge), 1×(Si-Ge), 4×(Si-Si)] picture. The 1D-Cluster re-assignment is independently secured by *ab initio* insight into the *frequency* of bond-stretching modes along prototype impurity motifs. These are taken as pseudo-linear so as to remain consistent with the LCA, upon which the 1D-Cluster scheme relies. Fair contour modeling of the SiGe Raman lineshapes is eventually achieved on this basis after *ab initio* calibration of the intrinsic Ge-Ge, Si-Ge and Si-Si Raman efficiencies, and proper weighting of such efficiencies by the fractions of related multi-oscillators in the crystal. The latter are likewise estimated within the LCA assuming a random Si↔Ge substitution.

More precisely, a predisposition of the Ge-Ge bond to exhibit a multi-mode Raman pattern, as evidenced by *ab initio*, seems to be impeded by parasitical disorder-induced effects. In fact a unique Raman mode is currently detected in the Ge-Ge spectral range, indicating a basic insensitivity of the Ge-Ge bond-stretching to its local environment. In contrast, significant fine structuring exists for both the Si-Si and Si-Ge bond-stretchings, revealing sensitivities to their $1^{st}$- and $2^{nd}$-neighbors 1D-environments, respectively. However, none of these is able to discriminate between all possible variants of such environments. Si-Si merely distinguishes between all-Ge (bottom branch) and other environments (top branch), i.e. including at least one Si atom. Regarding Si-Ge, the extreme Raman features refer to all-Ge (bottom branch) and all-Si (top branch) environments, the other (mixed) ones providing a common feature in-between (intermediate branch). A decomposition of the top Si-Ge branch soon after departing from the Ge-moderate limit, suggests some sensitivity beyond $2^{nd}$-neighbors. Nevertheless we are not able to identify the microstructure of such environments. For comparison the zincblende version of the 1D-Cluster scheme merely distinguishes between the two possible $1^{st}$ neighbor 1D-environments of a bond, as shown in this work.

Another major deviation between the SiGe and zincblende versions of the 1D-Cluster scheme, taking GaAsP as a reference, is concerned with an inversion of the order of the like phonon branches in each 1D-Cluster multiplet. This is attributed either to the considerable phonon dispersion of the Ge and Si crystals (Si-Si doublet), or to a basic difference between the diamond and zincblende lattice relaxations (Si-Ge triplet). The SiGe vs. GaAsP comparison is supported by *ab initio* bond length and BZC-phonons calculations using impurity motifs that are transposable from the zincblende structure to the diamond one.

Generally, this work constitutes the first extension of the Percolation/1D-Cluster scheme for the very basic understanding of the Raman spectra of mixed crystals beyond the current zincblende structure. We hope that it will stimulate further extension to alternative crystal structures, none of these being excluded in principle.